\documentclass[aps,prl,reprint,longbibliography,superscriptaddress]{revtex4-1}

 \usepackage{amsmath,bm}
 \usepackage{mathrsfs}
 \usepackage{amsfonts}
 \usepackage{graphicx}
 \usepackage{setspace} 
 \usepackage{graphicx}
 \usepackage{epstopdf}
 \usepackage{dcolumn}
 \usepackage{amsmath}
 \usepackage{epsfig}
 \usepackage{indentfirst}
 \usepackage{psfrag}
 \usepackage{subfigure}
 \usepackage{amssymb}
 \usepackage{color}
 \usepackage{units} 

 \usepackage{graphicx}
 \usepackage{dcolumn}
 \usepackage{bm}
 \usepackage{natbib}

\usepackage{physics}
\usepackage{dcolumn}
\usepackage{bm}
\usepackage{natbib}
\usepackage[backref=none,bookmarksnumbered=true,bookmarks=true,bookmarksopen=true,colorlinks=true,
citecolor=blue,linkcolor=blue,anchorcolor=green,urlcolor=blue,unicode=false]{hyperref}

\usepackage{ulem}[normalem] 

\def\black{\color{black}}

\makeatletter
\newcommand\colorsout[1]{\bgroup \markoverwith{\textcolor{#1}{\rule[0.5ex]{2pt}{0.4pt}}}\ULon}

\makeatother

\usepackage{xr-hyper}

\makeatletter
\newcommand*{\addFileDependency}[1]{
  \typeout{(#1)}
  \@addtofilelist{#1}
  \IfFileExists{#1}{}{\typeout{No file #1.}}
}
\makeatother

\usepackage{xr}
\graphicspath{ {Fig/} {Figsi/} }  

\begin{document}

\title{Tuning the Spin Interaction in Non-planar Organic Diradicals Through Mechanical Manipulation}

\author{Alessio Vegliante}
\affiliation{CIC nanoGUNE-BRTA, 20018 Donostia-San Sebasti\'an, Spain}

\author{Saleta Fernandez}
\affiliation{Centro Singular de Investigaci\'on en Qu\'imica Biol\'oxica e Materiais Moleculares (CiQUS) and Departamento de Qu\'imica Org\'anica, Universidade de Santiago de Compostela 15782-Santiago de Compostela (Spain)}

\author{Ricardo Ortiz}
\affiliation{Donostia International Physics Center (DIPC), 20018 Donostia-San Sebastián, Spain}

\author{Manuel Vilas-Varela}
\affiliation{Centro Singular de Investigaci\'on en Qu\'imica Biol\'oxica e Materiais Moleculares (CiQUS) and Departamento de Qu\'imica Org\'anica, Universidade de Santiago de Compostela 15782-Santiago de Compostela (Spain)}

\author{Thomas Baum}
\affiliation{Kavli Institute of Nanoscience, Delft University of Technology, 2628 Delft, The Netherlands} 

\author{Niklas Friedrich}
\affiliation{CIC nanoGUNE-BRTA, 20018 Donostia-San Sebasti\'an, Spain}

\author{Francisco Romero Lara}
\affiliation{CIC nanoGUNE-BRTA, 20018 Donostia-San Sebasti\'an, Spain}

\author{Andrea Aguirre}
\affiliation{Centro de Física de Materiales CSIC/UPV-EHU-Materials Physics Center, 20018 Donostia-San Sebasti\'an, Spain}

 \author{Katerina Vaxevani}
\affiliation{CIC nanoGUNE-BRTA, 20018 Donostia-San Sebasti\'an, Spain}
 
\author{Dongfei Wang}
\affiliation{CIC nanoGUNE-BRTA, 20018 Donostia-San Sebasti\'an, Spain}

\author{Carlos Garc\'ia}
\affiliation{Donostia International Physics Center (DIPC), 20018 Donostia-San Sebastián, Spain}

\author{Herre S. J. van der Zant}
\affiliation{Kavli Institute of Nanoscience, Delft University of Technology, 2628 Delft, The Netherlands}

\author{Thomas Frederiksen}
 \email{thomas_frederiksen@ehu.eus}
\affiliation{Donostia International Physics Center (DIPC), 20018 Donostia-San Sebastián, Spain}
\affiliation{IKERBASQUE, Basque Foundation for Science, 48013 Bilbao, Spain}

\author{Diego Peña}
 \email{diego.pena@usc.es}
\affiliation{Centro Singular de Investigaci\'on en Qu\'imica Biol\'oxica e Materiais Moleculares (CiQUS) and Departamento de Qu\'imica Org\'anica, Universidade de Santiago de Compostela 15782-Santiago de Compostela (Spain)}

\author{Jose Ignacio Pascual}
 \email{ji.pascual@nanogune.eu}
\affiliation{CIC nanoGUNE-BRTA, 20018 Donostia-San Sebasti\'an, Spain}
\affiliation{IKERBASQUE, Basque Foundation for Science, 48013 Bilbao, Spain}

\date{\today}

\newpage

\begin{abstract}
\setstretch{1.2}
Open-shell polycyclic aromatic hydrocarbons (PAHs) represent promising building blocks for carbon-based functional magnetic materials. Their magnetic properties stem from the presence of unpaired electrons localized in radical states of $\pi$ character. Consequently, these materials are inclined to exhibit spin delocalization, form extended collective states, and respond to the flexibility of the molecular backbones.   However, they are also highly reactive, requiring structural strategies to protect the radical states from reacting with the environment. Here, we demonstrate that the open-shell ground state of the diradical \textbf{2-OS} survives on a Au(111) substrate as a global singlet formed by two unpaired electrons with anti-parallel spins coupled through a conformational dependent interaction. The \textbf{2-OS} molecule is a “protected” derivative of the Chichibabin’s diradical, featuring a non-planar geometry that destabilizes the closed-shell quinoidal structure. Using scanning tunneling microscopy (STM), we localized the two interacting spins at the molecular edges, and detected an excited triplet state a few millielectronvolts above the singlet ground state. Mean-field Hubbard simulations reveal that the exchange coupling between the two spins strongly depends on the torsional angles between the different molecular moieties, suggesting the possibility of influencing the molecule's magnetic state through structural changes. This was demonstrated here using the STM tip to manipulate the molecular conformation, while simultaneously detecting changes in the spin excitation spectrum. Our work suggests the potential of these PAHs for a new class of all-carbon spin-crossover materials. 
\end{abstract}

\maketitle
Carbon-based molecular nanostructures can exhibit magnetic states associated with the stabilization of unpaired electrons in radical sites \cite{Oteyza2022}. Intrinsic $\pi$-magnetism has been widely observed in polycyclic aromatic hydrocarbons (PAHs) with an open-shell ground state, i.e., possessing one or more unpaired $\pi$ electrons \cite{Das2017, Li2019, Mishra2020, Mishra2020Nat}. Compared to traditional inorganic materials, the magnetism associated with these systems features longer spin coherence times and more delocalized magnetic moments, combined with a high degree of chemical and mechanical tunability \cite{Sugawara2009, Naber2007, Frisenda2015}.
In this context, organic diradicals are of fundamental interest for understanding magnetic interactions at the molecular scale and developing control strategies \cite{Rajca1994}. Overall, open-shell organic systems have the potential to become a new class of functional magnetic materials with applications in diverse technologies such as sensorics, non-linear optics, or spintronics \cite{Lehmann2009, Sanvito2011, Shil2015, Fukuda2015, Bogani2021, Dong2022}.

\begin{figure*} [th!]
   \includegraphics[width=0.96\textwidth]{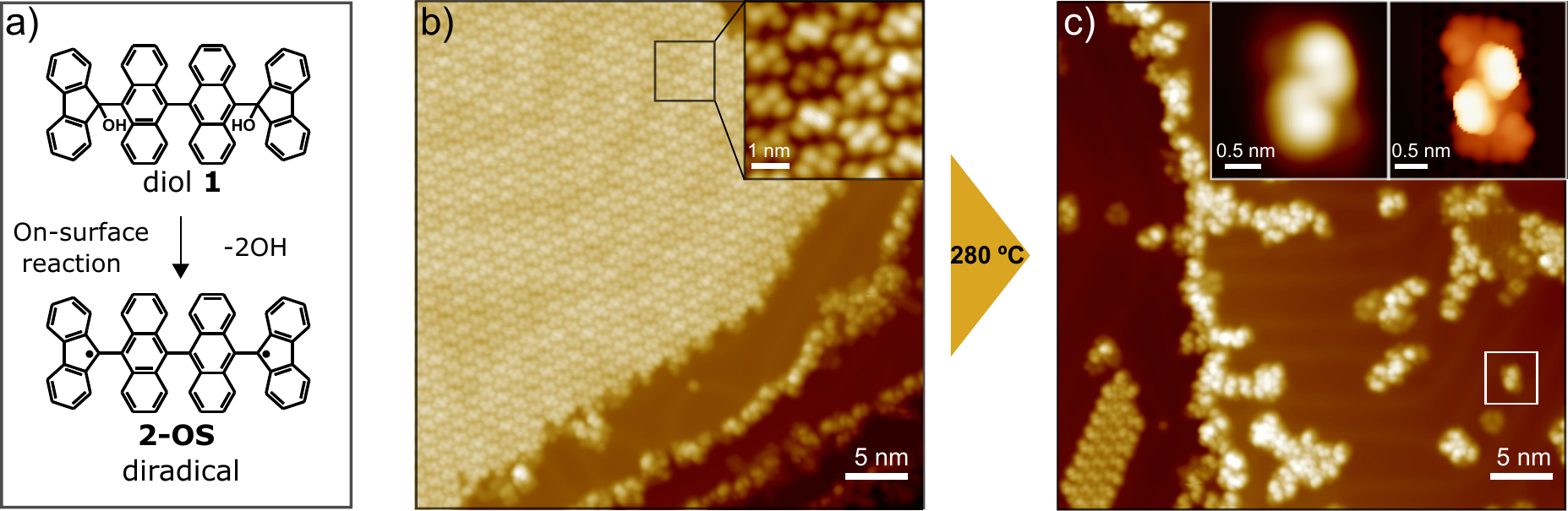}
	\caption{(a) Schematic representation of the generation of the \textbf{2-OS} diradical through dissociation of the OH groups from the diol \textbf{1} deposited on the Au(111) substrate. (b) Overview STM constant-current image displaying the typical close-packed domain formed by the molecular precursors when deposited on the Au(111) surface ($V=-1.25$ V; $I=46$ pA). In the inset, close-packed structure showing the arrangement and the shape of the individual molecules ($V=-1.25$ V; $I=30$ pA). (c) STM constant-current image recorded after annealing the sample at $280^{\circ}$C ($V=-1.25$ V; $I=30$ pA). In the inset: left, STM image of a single isolated molecule as found after the annealing ($V=-1.25$ V; $I=30$ pA); right, DFT charge density calculation of \textbf{2-OS} on Au(111). 
	}
\label{fig:Fig1}
\end{figure*}

A classical example of open-shell PAH is the Chichibabin's hydrocarbon, a molecule that has been extensively investigated for its large diradical character in the ground state \cite{Popp1978, Montgomery1986}. To circumvent its high reactivity, several derivatives have been prepared, one of the latest examples being a stable compound reported by Zeng et al. (\textbf{2-OS}, Figure \ref{fig:Fig1}a) \cite{Zeng2012}. The \textbf{2-OS} molecule consists of a central bisanthracene unit linked to two fluorenyl termini that accumulate the radical character of the system. The steric hindrance between fluorenyl and anthracene moieites determines a highly non-planar structure that is responsible for the protection of the radical centers, which remain localized over the fluorenyl sub-units. Owing to high stability of its diradical ground state and its tuneable non-planar structure,  \textbf{2-OS} is a suitable system to explore the relationship between magnetism and geometry at the single molecule scale.

A triplet ground state was reported for \textbf{2-OS} in solution, in agreement with DFT calculations of the magnetic state of the molecule in the gas phase \cite{Zeng2012}. Recently, \textbf{2-OS} was also studied in mechanically controlled break-junction devices \cite{Baum2022}, where spectroscopic features of either singlet and triplet ground state were detected, suggesting an interplay between the spin coupling and the precise structural conformation of the molecule. In order to unravel the nature of the intramolecular spin coupling in \textbf{2-OS}, here we study the magnetic properties of this system on a Au(111) surface at the single-molecule level by means of low-temperature scanning tunneling microscopy (STM) and spectroscopy (STS).
Our measurements, combined with mean-field Hubbard (MFH) calculations, reveal the influence of the molecular geometry on the magnetic coupling between the radical centers. Following these findings, we demonstrate the possibility of tuning the spin state of the molecule by modifying the arrangement of its constituent units through mechanical manipulation with the STM tip.

\section{Results and discussion}
\textbf{On-surface generation and characterization of 2-OS.} Evaporating di- and poly-radicals onto a substrate is known to be particularly challenging, as it easily results in fragmentation due to thermal instability \cite{Huang2017,Junghoefer2021}.
In order to get intact \textbf{2-OS} molecules on a Au(111) surface, we sublimated diol \textbf{1}, a stable molecular precursor containing a hydroxyl group (OH) capping each of the two radical centers at the fluorenyl termini, and subsequently induced an on-surface reaction to dissociate the OH groups and generate \textbf{2-OS}, as shown schematically in Figure \ref{fig:Fig1}a.

The diol \textbf{1} was evaporated onto a clean Au(111) surface under ultra-high vacuum (UHV) conditions. Extended close-packed molecular domains were found on the Au surface after the sublimation, as revealed by constant-current STM images (Figure \ref{fig:Fig1}b). A closer look into these structures shows that the constituent molecules appear partially planarized, and display four lobes: the two internal ones can be attributed to the anthracene units and the ones at the ends to the fluorenyl termini. Differential conductance ($dI/dV$) spectra acquired on the molecules inside the close-packed structures show no fingerprints of magnetism (Figure S7). This is expected as a consequence of the presence of the OH groups of compound \textbf{1}, that are not detached upon deposition onto the surface and still passivate the radical centers, preserving the closed-shell structure of the precursor.

The sample was thus annealed with the aim of inducing the C-OH cleavage, in a similar manner to the deoxygenation reaction previously described for epoxyacene derivatives \cite{Eisenhut2020}. Figure \ref{fig:Fig1}c shows an overview after annealing at $280^{\circ}$C: some close-packed domains of reduced dimensions can still be seen, alongside single isolated molecules, chains and small molecular clusters of several shapes. Individual molecules appear in constant-current STM images as shown in the inset in Figure \ref{fig:Fig1}c: they feature two internal brighter lobes, that can be attributed to the anthracene moieties, and darker external features, corresponding to the fluorenyl termini. A more detailed elucidation of the molecular structure with bond resolution is hindered by the non-planarity of the system. However, our DFT simulations of the charge density of \textbf{2-OS} on the Au(111) surface (also reported in Figure \ref{fig:Fig1}c) is in good agreement with the main features found in the STM topography, confirming our identification of the individual molecules. 

\begin{figure*}[ht]
    \includegraphics[width=\textwidth]{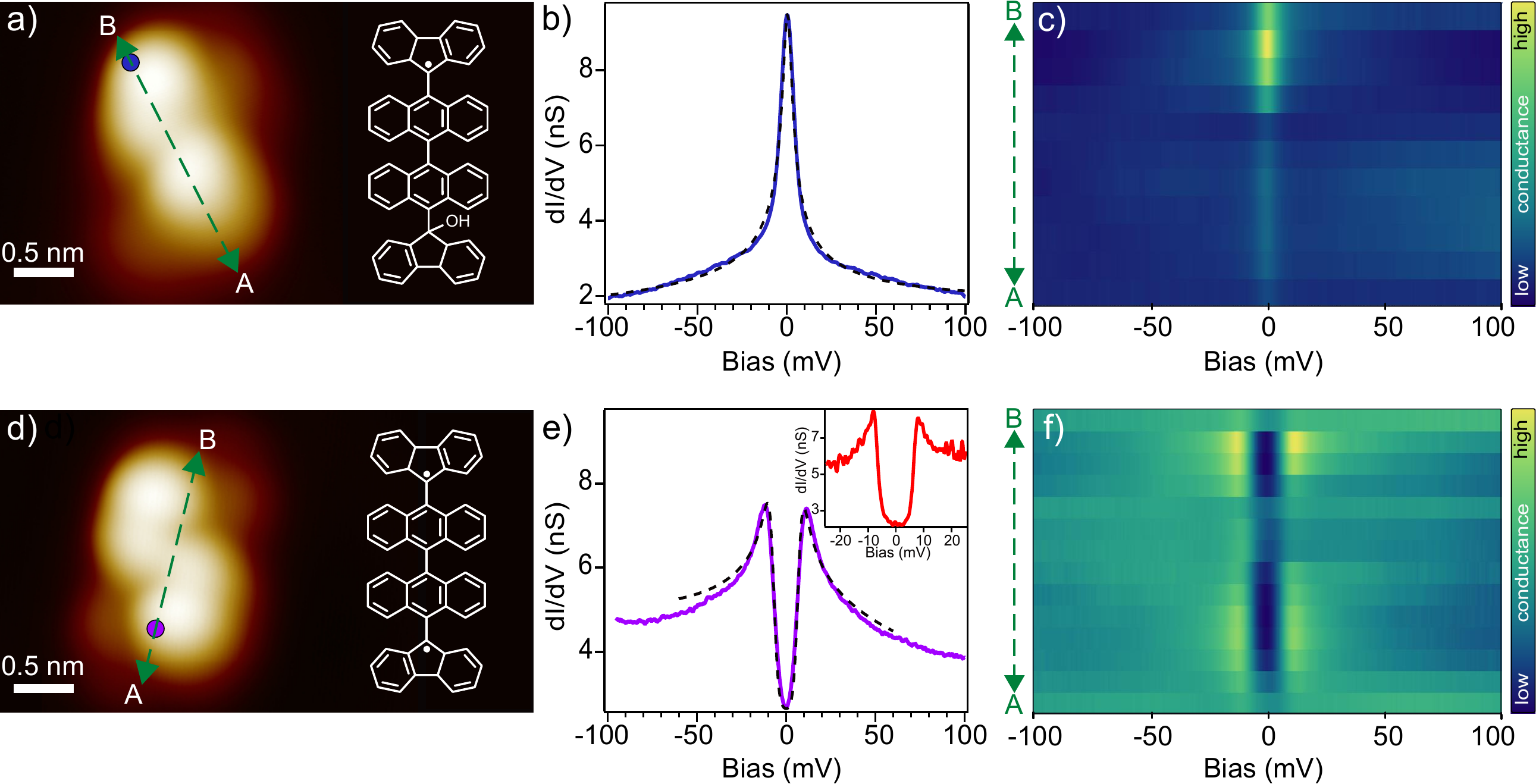}
        \caption{(a,d) STM constant-current images ($V=-1.25$ V; $I=30$ pA) of single molecules as found after the on-surface reaction and the corresponding molecular structures, in the two configurations: (a) with a residual OH group, corresponding to a monoradical, and (d) with no OH groups, corresponding to the \textbf{2-OS} diradical. (b) $dI/dV$ spectrum taken on the molecule in (a) at the position indicated by the blue circle, displaying a zero-bias resonance, that can be fitted with a Frota function (black dashed line) \cite{Frota1992}. (c) $dI/dV$ linescan measured across the molecule in (a) in the direction indicated by the arrow ($V=-200$ mV, $I_{set}=500$ pA, $V_{mod}=2$ mV). (e) $dI/dV$ spectrum measured on the molecule in (d) at the position indicated by the purple circle, showing spin excitation steps. The black dashed line represents fits to the data using the perturbative model by Ternes \cite{Ternes2015}, from which an antiferromagnetic exchange $J=7.3$ meV is obtained. The inset shows a spectrum measured at $T=1.3$ K, emphasizing the IET gap ($V$=30 mV, $I_{set}=200$ pA, $V_{mod}=0.8$ mV). (f) $dI/dV$ line spectra measured across the molecule in (d) in the direction indicated by the arrow ($V=-100$ mV, $I_{set}=500$ pA, $V_{mod}=2$ mV).}
    \label{fig:Fig2} 
\end{figure*}

An alternative approach for the on-surface formation of \textbf{2-OS} consists in inducing the C-OH cleavage by means of a voltage pulse, obtained by placing the tip on top of a molecule and raising the sample positive bias above $1.5$ V. As shown in the supplementary  information (Figure S8), this procedure allows to remove one or more OH groups in the same molecule or in a neighboring molecule within the assembled domains and molecular clusters, but does not provide isolated molecules over the substrate. Therefore, our further analysis is based on single molecules obtained by annealing the substrate. 

Two distinct molecular structures are generally found after the annealing step, corresponding to one or both OH groups detached following the on-surface reaction (Figure \ref{fig:Fig2}a,d). The STM constant-current images appear roughly the same in the two cases; nevertheless, scanning tunneling spectroscopy allows the identification of two distinct spin states, that can be related to the two configurations with different numbers of OH groups left. 

In a fraction of molecules (around 25\% of the single molecules that have been investigated, Figure \ref{fig:Fig2}a), the $dI/dV$ spectra display a pronounced zero-bias peak, as depicted in Figure \ref{fig:Fig2}b. This feature is well reproduced by a Frota function with $HWHM=6.1$ meV, and can be attributed to a Kondo resonance, arising from the screening of a localized spin $S=1/2$ by the conduction electrons of the metal substrate \cite{Kondo1964, Ternes2008, Li2019, Li2020, Frota1992}. The $dI/dV$ stacked plot taken along an axis of the molecule (Figure \ref{fig:Fig2}c) shows that the Kondo resonance is not spatially homogeneous but significantly more intense in one half of the molecule. We relate this spin state to a monoradical structure in which one OH group persists after the annealing and therefore conclude that the Kondo resonance is associated to the single unpaired electron recovered from the partial OH dissociation.

However, most of the individual molecules found after annealing (Figure \ref{fig:Fig2}d) display a distinct low-energy feature consisting of a narrow gap centered at zero-bias, followed by two sharp $dI/dV$ peaks at $\pm 11$ meV, as depicted in Figure \ref{fig:Fig2}e. The $dI/dV$ stacked plot in Figure \ref{fig:Fig2}f reveals that these features appear distributed all over the molecule, weaker over the center of the molecule but with higher amplitude towards the fluorenyl end-groups. Based on the symmetric position of the $dI/dV$ peaks, we attribute these features to an inelastic excitation of the two exchange-coupled spins at the radical sites \cite{Hirjibehedin2006, Li2019, Hieulle2021, Mishra2020, Ortiz2020, Hieulle2023}, confirming that in this case both OH groups have been detached, activating the \textbf{2-OS} diradical. 

We note that $dI/dV$ spectra exhibit a higher-bias characteristic fall-off, resembling a Kondo resonance superimposed on the gapped spectrum \cite{Li2019,Hieulle2023}. Furthermore, such Kondo-like feature is absent at zero bias (see STS spectrum recorded at 1.3K in inset in Figure \ref{fig:Fig2}e). This indicates that the Kondo-like fluctuations are enabled by inelastic electrons tunneling through the excited state \cite{Paaske2006,Ternes2008} but are absent in the ground state. The spectral shape can be interpreted as caused by the inelastic excitation of a singlet (total spin $S=0$) ground state into a triplet ($S=1$) excited state. A model of two antiferromagnetically coupled $1/2$ spins \cite{Ternes2015} (dashed line in Figure \ref{fig:Fig2}e) reproduces well the spectral features, revealing an exchange interaction between the flourenyl moieties of $J\sim~7.3$ meV. Therefore,  we conclude that \textbf{2-OS} on a Au(111) surface exhibits a singlet ground state, excluding the previously observed triplet ground state \cite{Zeng2012}, since this would result in a very different spectral lineshape.
 
\begin{figure}[ht!]
   \includegraphics[width=\columnwidth]{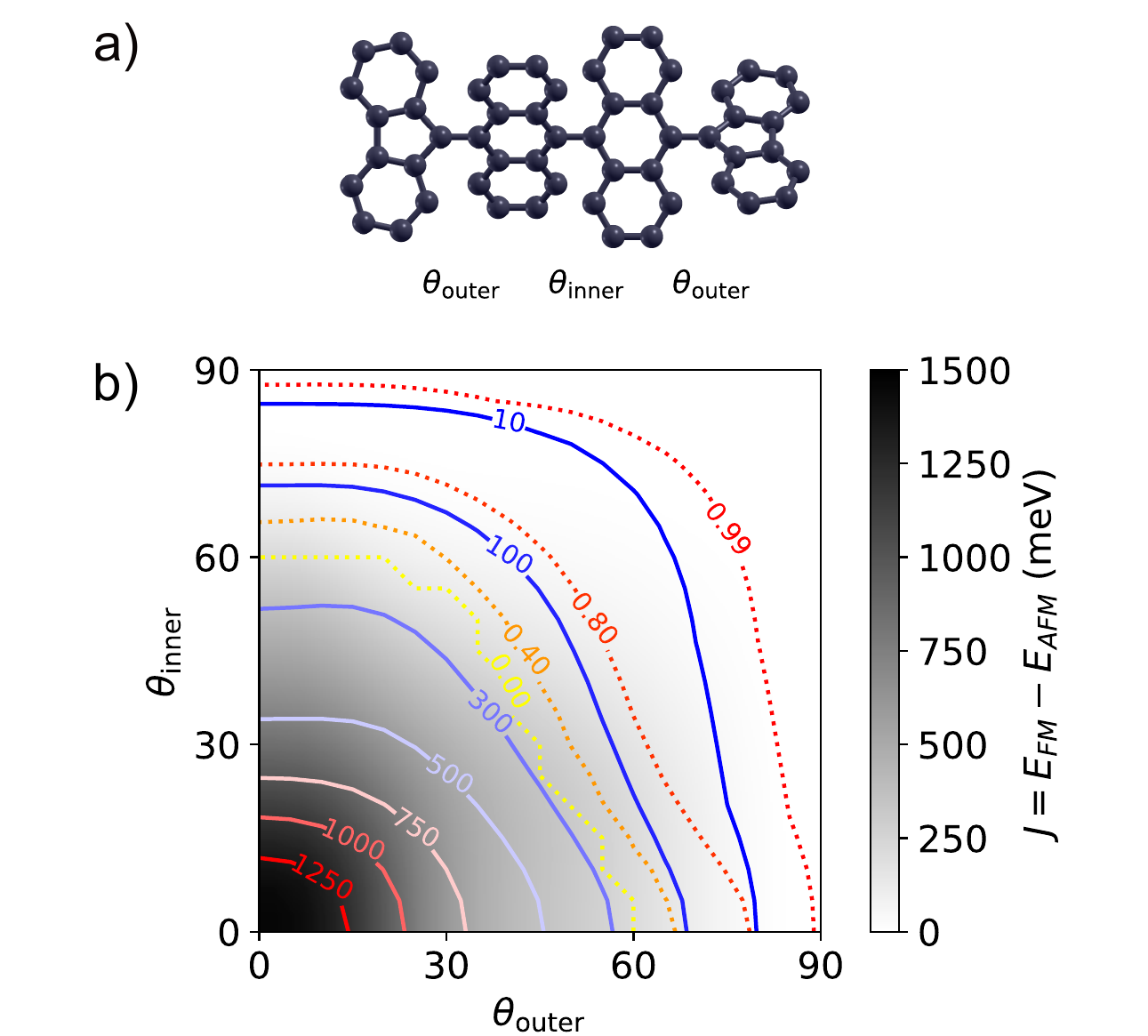}
	\caption{
	 {TB-MFH calculations for the exchange coupling as a function of torsional angles in \textbf{2-OS}}.
  (a) Structural model for the \textbf{2-OS} carbon backbone with three torsional angles.
  For simplicity, we fix the two outer angles $\theta_\mathrm{outer}$ to be identical.
  (b) Using MFH with $U=3$ eV \cite{dipc_hubbard}, we solve for the energy difference $J$ between the ferromagnetic (FM) and antiferromagnetic (AFM) solutions
  as a function of the two torsional angles $\theta_\mathrm{outer}$ and $\theta_\mathrm{inner}$.
  The full-line contours represent constant-$J$ while the dashed contours represent the local polarization per radical (sum of the spin density over sites on one half of the molecule) in the AFM solution.
  The AFM solution is always the one with the lowest electronic energy, but
  $J$ approaches zero whenever one of the two angles reaches $90^\circ$.
  When $J\gtrsim 200$ meV the AFM solution corresponds to the closed-shell singlet (no local polarization).
}
\label{fig:Fig3}
\end{figure}

\textbf{Theoretical calculations.} To explain the origin of the antiferromagnetic coupling found on the Au substrate,
we performed  tight-binding mean-field Hubbard (TB-MFH) simulations of the magnetic ground state of \textbf{2-OS} in different geometries. Specifically, we calculated the energy  of the antiferromagnetic (AFM, the singlet case) spin arrangement with respect to the ferromagnetic (FM, the triplet solution) case, i.e. $J=E_{FM}-E_{AFM}$, as a function of the dihedral  angles  $\theta_\mathrm{inner}$ (between the central anthracenes) and $\theta_\mathrm{outer}$ (between each anthracene and the outer fluorenyl). For simplicity, we assumed the two outer angles to be identical and the individual anthracenes and fluorenyls units to be planar. We considered a single $p$-orbital per carbon site, locally perpendicular to the backbone unit and constructed the corresponding TB Hamiltonian with \textsc{sisl} \cite{zerothi_sisl} using the Slater-Koster parametrization of \citet{RePh.18.Modifiedspinorbit} (see Supplementary Information Section 3.1).

The variation of $J$ as a function of the two torsional angles is depicted in Figure \ref{fig:Fig3}b. The AFM solution is always the lowest in energy 
for any combination of $\theta_\mathrm{inner}$ and $\theta_\mathrm{outer}$. However, we observe a clear trend in the evolution of $J$ with the conformation: its value is close to 0 when any of the torsional angles approach $90^\circ$, but it progressively increases as the angles reduce. This means that the planarization of the molecular structure (i.e. the decrease of the angles between the units) stabilizes the singlet solution, unveiling an increasingly higher AFM exchange coupling. According to the TB-MFH results, the effect is drastic: the open-shell singlet vanishes for angles of just $60^\circ$, transforming itself into a (spin non-polarized) closed-shell solution. 

The experimental observation of a singlet ground state for \textbf{2-OS} on Au(111) is in agreement with the trend obtained in the MFH simulations. The interaction with the flat metal substrate is expected to induce a partial planarization of the molecular units (see the DFT calculation of the adsorption geometry in Supplementary Information Section 3.2), contributing to the stabilization of the AFM order. 
From TB-MFH results, the FM solution is always less energetically favorable than the AFM one, even for high values of the torsional angles.
This apparently contradicts results from previous DFT calculations, which found a triplet ground state in the gas phase, with a triplet-singlet gap of $26$~meV \cite{Zeng2012}. This discrepancy is explained by considering that the value reported by Zeng et al. \cite{Zeng2012} takes into account also vibrational effects due to temperature. Including these in our DFT simulations, we reproduced the higher stability of the triplet ground state for a gas-phase relaxed molecule (i.e. with both angles $\theta_\mathrm{i} \sim$ 90$^{\circ}$) and at room temperature (shown in Supplementary Information, Section 3.3). At the lower temperature of our experiment, however, vibrational effects do not play a significant role, and DFT qualitatively reproduces the results of the TB-MFH calculations. 

\begin{figure*}[t]
   \includegraphics[width=\textwidth]{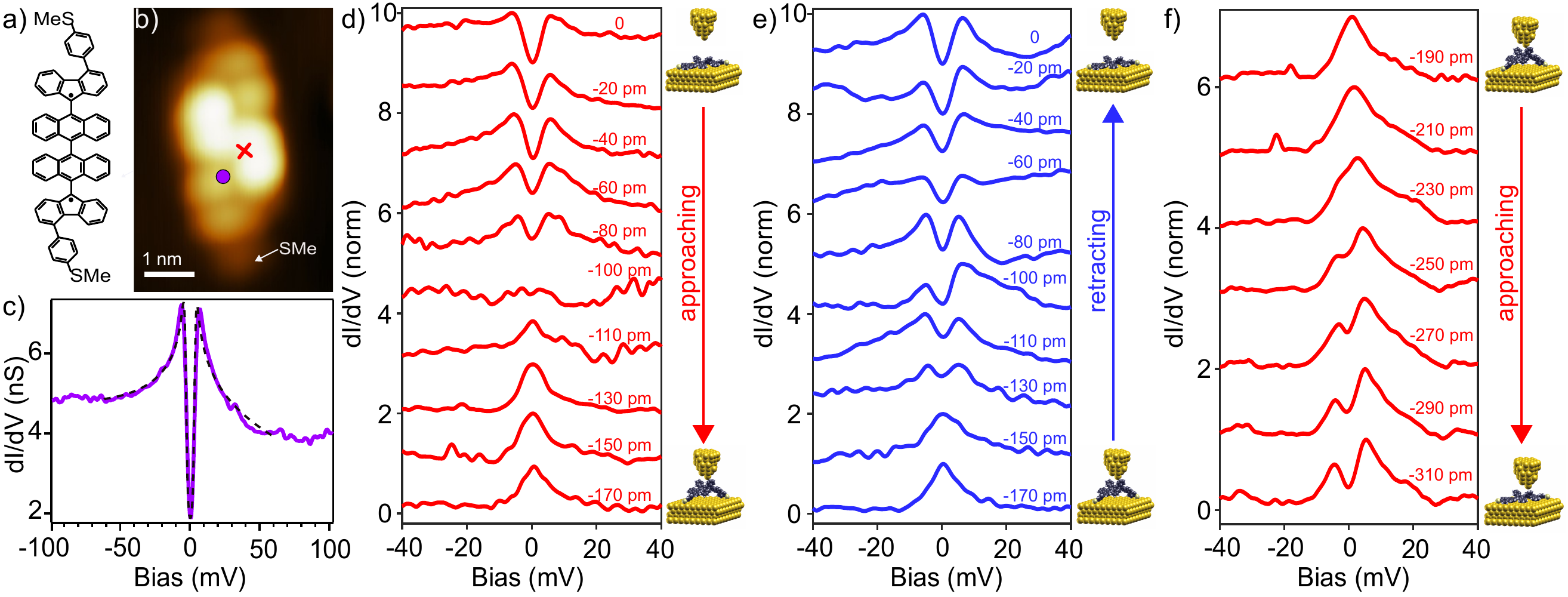}
	\caption{(a) Chemical model of \textbf{SMe-2OS}, the sulfide substituted analogue of \textbf{2-OS} synthesized for the manipulation experiment. (b) STM  image ($V=-1.25$ V; $I=30$ pA) of \textbf{SMe-2OS}, as obtained after the on-surface generation of the diradical, reported in the Supplementary Information. (c) $dI/dV$ spectrum measured on the molecule at the position indicated by the purple circle in (b), displaying the same IET feature as \textbf{2-OS}. The black dashed line is a fit to the data using the perturbative model by Ternes \cite{Ternes2015}, revealing an antiferromagnetic exchange $J=3$ meV. (d-f) Mechanical manipulation of the spin state of \textbf{SMe-2OS}. The STM tip is stabilized on the molecule in the position marked by the red cross in (b) at a starting height (here indicated as 0) determined by the offset parameters $I=200$ pA and $V=50$ mV. Then, after opening the feedback loop, $dI/dV$ spectra are measured at different tip heights, in three ranges: (d) while approaching the tip down to 170pm with respect to the starting position; (e) when retracting back towards the starting position; (f) while continuing approaching to lower tip-sample distances than the ones represented in range (d). On the right of each of the three sets of $dI/dV$ measurements a model represents the process of partially lifting/releasing/pushing the molecule. Spectroscopy parameters: $V_{mod}=2$ mV, $I_{set}=200$ pA.}
\label{fig:Fig4}
\end{figure*}

In any scenario, all simulation tools agree that decreasing the torsional angle of the molecular subunits stabilizes the singlet state. As shown in Figure S11 and S12, this can be explained by the increase in hopping elements across the dihedral angles. These matrix elements primarily enhance kinetic exchange mechanisms, which induce AFM ordering in \textbf{2-OS}, while direct Hund's like exchange is expected to be negligible \cite{Jacob2022,Yu2023}. Following these theoretical results, the controlled modification of the spin interactions should be achievable by tuning the molecular configuration. 
\black

\textbf{Mechanical manipulation of the spin-spin coupling.}
Motivated by the theoretical predictions indicating a correlation between the geometry of \textbf{2-OS} and the exchange coupling $J$, we explored the possibility of modifying its spin excitation gap and  magnetic ground state by controllably manipulating the molecular conformation  using the STM tip. We note that the  non-planar structure of \textbf{2-OS} arises from hindrance among different conjugated units, imposing the persistance of a dihedral angle even on the metal substrate. Consequently, we anticipated that approaching the STM tip towards the central part of  the molecule would exert attractive or repulsive forces over these submolecular units, forcing variations of their dihedral angle \cite{Heinrich2018, Vaxevani2022}.

However, due to the weak interaction with the surface, \textbf{2-OS} molecules were abruptly modified, and frequently displaced towards the STM tip in response to  attractive forces (see Figure S9). To overcome this issue, we designed an extended version of the diradical, denoted \textbf{SMe-2OS} (Figure \ref{fig:Fig4}a), wherein methylthiophenyl moieties were strategically incorporated at each fluorenyl sub-unit. These moieties were chosen since sulfur groups are widely used for anchoring molecules to gold electrodes \cite{Heersche2006,Naghibi2022,Lokamani2023}.

Similar to the \textbf{2-OS} base molecule, we generated the \textbf{SMe-2OS} diradical on the Au(111) surface by depositing the corresponding closed-shell precursor containing OH protecting groups at the two radical sites (see structure of diol \textbf{10} in Figure S1). In this case, we activated the OH decapping by applying controlled bias pulses over the molecular center, as illustrated in Figure S10.

The STM constant-current image of the resulting  \textbf{SMe-2OS} molecule, shown in Figure \ref{fig:Fig4}b, reveals similar bulky features like in \textbf{2-OS} species but with the addition of two smaller bright lobes at each end, attributed to the methylthiophenyl end groups. 

The $dI/dV$ spectra measured on \textbf{SMe-2OS} (Figure \ref{fig:Fig4}c) reproduce similar features as in \textbf{2-OS}, namely,  IET  onsets attributed to singlet-triplet spin excitation, accompanied by a $dI/dV$ decrease characteristic of Kondo fluctuations in the excited state. Therefore, apart from a slightly reduced value of the singlet-triplet gap  $\sim 3$~meV in \textbf{SMe-2OS}), the addition of the edge groups preserves the magnetic ground state of \textbf{2-OS}. 

To study the evolution of spin interactions in \textbf{SMe-2OS} as a function of structural changes, we stabilized the STM tip above the center of the molecule (red cross in Figure \ref{fig:Fig4}b) and performed $dI/dV$ measurements while the STM tip was approached to or retracted from the molecule in steps of 10 or 20~pm. 

Figure \ref{fig:Fig4}d displays spectra measured while approaching the tip 170~pm from the starting position. The width and depth of the excitation gap decreases along the first approaching steps until  collapsing into a zero-bias peak at 100~pm. This change can be interpreted as a transition to a new spin state of \textbf{SMe-2OS} caused by the structural modification induced by the tip. Approaching the tip towards the molecule is known to induce first attractive forces \cite{Farinacci2018, Trivini2023} that here produce a rearrangement of the internal anthracene units into a less planar conformation, as shown in the schematic representation of Figure \ref{fig:Fig4}d.  As suggested by the theory, an increase in one or both the torsional angles results in the reduction of the intramolecular spin-spin coupling. Therefore, the zero-bias peak observed in the final steps of the approach can be attributed to a Kondo-like feature, emerging when the two spins become non-interacting.

The mechanical manipulation of the spin state of the diradical is reversible {(Figure \ref{fig:Fig4}e). Retracting the tip back to the starting position restores the initial inelastic spectral features, thus indicating that the molecule is brought back into the original conformation stabilized by the substrate. The reversibility of this process demonstrates that the change in the spectra is due to mechanical modifications rather than to the formation of a tip-molecule bond (and consequent quenching of a spin). 

When the STM tip is instead moved further towards the molecule from the closest point in Figure \ref{fig:Fig4}d, the antiparallel spin configuration is also restored.
Figure \ref{fig:Fig4}f shows $dI/dV$ spectra measured while approaching the tip from 190~pm to 310~pm from the starting position in Figure~\ref{fig:Fig4}d. In this range, the Kondo peak gradually develops a gap with tip approach attributed to the reactivation of an inelastic spin excitation. This indicates that repulsive tip-molecule forces start to play a role and the molecule is pushed back towards the substrate, progressively restoring the torsional angles between the central anthracenes and recovering the antiparallel spin interaction. 
Curiously, this gap-reopening is smooth, in contrast to the first range of the approach (Figure \ref{fig:Fig4}d), where a sudden gap closing took place. This suggests that in the range of pushing, the dihedral angle can be tuned gradually, while during the initial pulling regime, the attractive forces acting at larger distances induce a more sudden structural reorganization.
These results demosntrate that the spin-spin coupling in \textbf{2-OS} can be tuned by means of mechanical manipulation in a reversible way, confirming the theoretical prediction of a strong interplay between precise structural conformation and magnetic state. 

\section{Conclusions}
In summary, we have demonstrated that the Chichibabin's hydrocarbon \textbf{2-OS} retains its open-shell character on a Au(111) surface with a singlet ground state  consisting of two antiparallel aligned  spins. Through the cleavage of protecting groups of the deposited diol precursors followed by spatially resolved scanning tunnelling spectroscopy measurements we resolved singlet-triplet excitations at the energy $J\sim~7.3$~meV, thus quantifying their exchange coupling strength.  Theoretical MFH simulations found that the substrate-induced partial planarization of \textbf{2-OS}  favours the antiparallel coupling of the unpaired $\pi$ electrons, while increasing the torsional angle between the molecular moieties progressively reduces the spin exchange coupling. The ability to tailor magnetic interactions by torsional forces is a promising avenue for the realization of molecule-based sensors. Therefore, we demonstrated this possibility using a \textbf{2-OS} molecule functionalized with anchoring end groups, which fix the molecule to the metal while inducing structural modifications with the STM tip. We observed that the spin-spin interaction decreases when the molecule is partially lifted from the substrate (thus acquiring a more orthogonal conformation), and that the initial singlet state can be recovered when bringing it back to a more planar arrangement.  
The robust singlet ground state found at low temperature contrasts with the triplet ground state found at room temperature \cite{Zeng2012}. We attributed this temperature dependence to contributions of thermally excited vibrational deformations. Together with the reported high sensitivity of the magnetic ground state of these Chichibabin's hydrocarbons to mechanical interactions, our results suggest a strong potential of \textbf{2-OS} as all-organic spin-cross over material.

\section*{Methods}
The (111) surface of a single Au crystal was cleaned by several cycles of sputtering with Ne gas and successive annealing at $T=600$°C under ultrahigh vacuum (UHV) conditions. The precursors of \textbf{2-OS} and \textbf{SMe-2OS} were prepared in solution following the protocol described in the Supplementary Information (Section 1) and evaporated onto the Au(111) surface held at room temperature. The sublimation of intact precursor molecules can be achieved via fast thermal heating of a silicon wafer loaded with grains of the compound, as well as through a standard Knudsen cell. All the measurements were performed in a custom-made low-temperature STM at $5$~K in UHV, with the exception of the $dI/dV$ spectrum in the inset of Figure \ref{fig:Fig2}e, measured in a commercial Joule-Thompson (JT) STM with a base temperature of 1.3~K.

Differential conductance spectra were recorded using a lock-in amplifier with frequency $f=867.9$ Hz (Figure \ref{fig:Fig2}) and $f=944$ Hz (Figure \ref{fig:Fig4}), with the modulation amplitude and current parameters indicated in the text. 

The distance-dependent $dI/dV$ measurements reported in Figure \ref{fig:Fig4}d-f were performed by stabilizing the tip on top of a molecule with the setpoint parameters indicated in the text and using a custom-built interface to control the tip movement z and simultaneously record the tip movement, the current, the bias and the lock-in signal. The spectra displayed in Figure \ref{fig:Fig4}d-f were obtained after subtracting a 3rd order polynomial background adjusted to each spectrum and then normalizing the conductance between the maximum and minimum value, according to the formula $G_{norm}(V) = (G(V)–G_{min})/(G_{max}–G_{min})$, where $G(V)$ is the differential conductance at bias $V$, and $G_{max}$ and $G_{min}$ the maximum and minimum value of $G(V)$ respectively.

All STM images and $dI/dV$ spectra were performed with gold-coated tungsten tips. The Figures representing experimental data were prepared using WSxM and SpectraFox softwares \cite{Horcas2007, Ruby2016}.
The MFH simulations in Figure \ref{fig:Fig3} and the DFT calculations were realized as explained in the Supplementary Information (Section 3) using the software packages \textsc{hubbard} \cite{dipc_hubbard}, \textsc{Gaussian} \cite{g16} and \textsc{Siesta} \cite{SoArGa.02.SIESTAmethodab}.

\section*{Acknowledgments}

The authors gratefully acknowledge financial support from the Spanish \\ MCIN/AEI/10.13039/501100011033 and the European Regional Development Fund (ERDF) through grants PID2022-140845OB-C61, PID2022-140845OB-C62, PID2020-115406GB-I00, and the Maria de Maeztu Units of Excellence Program CEX2020-001038-M,  from the European Union (EU) through the FET-Open project SPRING (863098), the ERC Synergy Grant MolDAM (951519), the ERC-AdG CONSPIRA (101097693),  and from the Xunta de Galicia (Centro singular de investigación de Galicia accreditation 2019-2022, ED431G 2019/03 and Oportunius Program). F.R.-L. acknowledges funding by the Spanish Ministerio de Educación y Formación Profesional through the PhD scholarship No. FPU20/03305.

\bibliographystyle{apsrev4-1} 
 

 \providecommand{\latin}[1]{#1}
\makeatletter
\providecommand{\doi}
  {\begingroup\let\do\@makeother\dospecials
  \catcode`\{=1 \catcode`\}=2 \doi@aux}
\providecommand{\doi@aux}[1]{\endgroup\texttt{#1}}
\makeatother
\providecommand*\mcitethebibliography{\thebibliography}
\csname @ifundefined\endcsname{endmcitethebibliography}
  {\let\endmcitethebibliography\endthebibliography}{}

\end{document}


\title{\textit{Supplementary Information}:\\Tuning the Spin Interaction in Non-planar Organic Diradicals Through Mechanical Manipulation} 

\author{Alessio Vegliante}
\affiliation{CIC nanoGUNE-BRTA, 20018 Donostia-San Sebasti\'an, Spain}

\author{Saleta Fernandez}
\affiliation{Centro Singular de Investigaci\'on en Qu\'imica Biol\'oxica e Materiais Moleculares (CiQUS) and Departamento de Qu\'imica Org\'anica, Universidade de Santiago de Compostela 15782-Santiago de Compostela (Spain)}

\author{Ricardo Ortiz}
\affiliation{Donostia International Physics Center (DIPC), 20018 Donostia-San Sebastián, Spain}

\author{Manuel Vilas-Varela}
\affiliation{Centro Singular de Investigaci\'on en Qu\'imica Biol\'oxica e Materiais Moleculares (CiQUS) and Departamento de Qu\'imica Org\'anica, Universidade de Santiago de Compostela 15782-Santiago de Compostela (Spain)}

\author{Thomas Baum}
\affiliation{Kavli Institute of Nanoscience, Delft University of Technology, 2628 Delft, The Netherlands} 

\author{Niklas Friedrich}
\affiliation{CIC nanoGUNE-BRTA, 20018 Donostia-San Sebasti\'an, Spain}

\author{Francisco Romero Lara}
\affiliation{CIC nanoGUNE-BRTA, 20018 Donostia-San Sebasti\'an, Spain}

\author{Andrea Aguirre}
\affiliation{Centro de Física de Materiales CSIC/UPV-EHU-Materials Physics Center, 20018 Donostia-San Sebasti\'an, Spain}

 \author{Katerina Vaxevani}
\affiliation{CIC nanoGUNE-BRTA, 20018 Donostia-San Sebasti\'an, Spain}
 
\author{Dongfei Wang}
\affiliation{CIC nanoGUNE-BRTA, 20018 Donostia-San Sebasti\'an, Spain}

\author{Carlos Garc\'ia}
\affiliation{Donostia International Physics Center (DIPC), 20018 Donostia-San Sebastián, Spain}

\author{Herre S. J. van der Zant}
\affiliation{Kavli Institute of Nanoscience, Delft University of Technology, 2628 Delft, The Netherlands}

\author{Thomas Frederiksen}
\affiliation{Donostia International Physics Center (DIPC), 20018 Donostia-San Sebastián, Spain}
\affiliation{IKERBASQUE, Basque Foundation for Science, 48013 Bilbao, Spain}

\author{Diego Peña}
\affiliation{Centro Singular de Investigaci\'on en Qu\'imica Biol\'oxica e Materiais Moleculares (CiQUS) and Departamento de Qu\'imica Org\'anica, Universidade de Santiago de Compostela 15782-Santiago de Compostela (Spain)}

\author{Jose Ignacio Pascual}
\affiliation{CIC nanoGUNE-BRTA, 20018 Donostia-San Sebasti\'an, Spain}
\affiliation{IKERBASQUE, Basque Foundation for Science, 48013 Bilbao, Spain}

\renewcommand{\abstractname}{\vspace{1cm} }	
\begin{abstract}
\baselineskip12pt
\end{abstract}	
\baselineskip14pt
\maketitle 


\baselineskip15pt
\section{Synthetic details}
\label{sec:chemistry}

\subsection{General methods for the synthesis in solution}
All the reactions were carried out under argon using oven-dried glassware. \ce{CH2Cl2} and tetrahydrofuran (THF) were dried using a MBraun SPS-800 Solvent Purification System. Finely powdered \ce{SnCl2} was purchased from Sigma-Aldrich, weighted and stored in a glove-box. Other commercial reagents were purchased from ABCR GmbH, Sigma-Aldrich or Acros Organics, and were used without further purification. Deuterated solvents were purchased from Acros Organics. TLC was performed on Merck silica gel 60 F254 and chromatograms were visualized with UV light (254 and 365 nm) and/or stained with Hanessian’s stain. Column chromatography was performed on Merck silica gel 60 (ASTM 230-400 mesh). \ce{^1H} and \ce{^13C} NMR spectra were recorded at 300 and 75 MHz (Varian Mercury-300 instrument) or 500 and 125 MHz (Varian Inova 500 or Bruker 500) respectively. APCI high resolution mass spectra were obtained on a Bruker Microtof. The synthesis of precursor \textbf{1} was performed following a known procedure \cite{Zeng2012}, while precursor \textbf{10} was obtained by a similar protocol shown in Scheme 1.

\begin{figure}[hbt]
    \includegraphics[width=0.9\columnwidth]{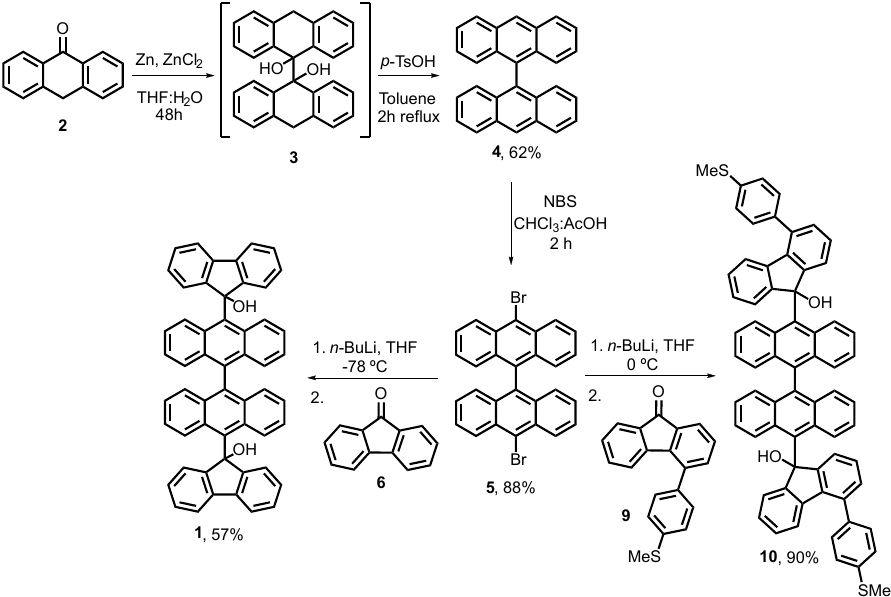}
    \caption{Scheme 1. Synthesis of molecular precursors \textbf{1} and \textbf{10}.}
    \label{fig:precursors}
\end{figure}

\newpage
\subsection{Synthesis of 9,9’-bianthracene (4)}
\begin{figure}[hbt]
    \includegraphics[width=0.6\columnwidth]{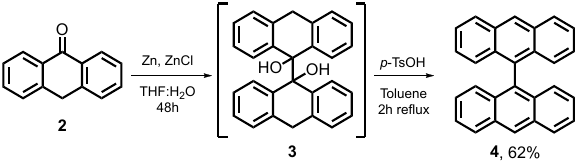}
\end{figure}	
9-Anthrone (\textbf{2}, 2.0 g, 10.3 mmol) was added over a suspension of Zn (3.34 g, 52 mmol) and \ce{ZnCl2} (2.8 g, 20.6 mmol) in THF:H2O (7: 3, 20mL) and the resulting mixture was stirred at room temperature during 48 h. Then, \ce{CH2Cl2} (20 mL) and \ce{H2O} (20 mL) were added, the phases were separated, and the aqueous phase was extracted with \ce{CH2Cl2} (2x10 mL). The combined organic phases were dried over \ce{MgSO4}, filtered, and concentrated under reduced pressure. The resulting residue was dissolved in toluene (15 mL) and added a catalytic amount of \ce{\textit{p}-TsOH}, the mixture was refluxed for 2h. The solvent was evaporated, and the residue obtained was purified by column chromatography (\ce{SiO2}; hexane: \ce{CH2Cl2} 4: 1) yielding the product \textbf{4} (1.13 g, 62\%) as a greenish solid \cite{Tanaka1990}.

\ce{^1H} NMR (300 MHz, \ce{CDCl3}) $\delta$: 8.68 (s, 1H), 8.17-8.14 (d, \textit{J} = 8.6 Hz, 2H), 7.47-7.42 (t, \textit{J} = 7.5 Hz, 2H), 7.17-7.06 (dt, \textit{J} = 16.2, 8.7 Hz, 4H) ppm.

\subsection{Synthesis of 10,10’-bromo-9,9’-bianthracene (5)}
\begin{figure}[hbt]
    \includegraphics[width=0.4\columnwidth]{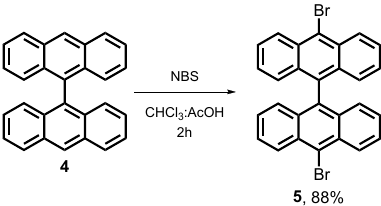}
\end{figure}
A mixture of compound \textbf{4} (1.13 g, 3.2 mmol) and NBS (1.42g, 7.97 mmol) in \ce{CHCl3}:\ce{CH3CO2H}(6:1) was refluxed for 16 h. The mixture was then allowed to cool down to room temperature, \ce{CHCl3} (40 mL) and \ce{H2O} (40 mL) were added, the phases were separated, and the aqueous phase was extracted with \ce{CHCl3} (2x50 mL). The combined organic phases were dried over \ce{MgSO4}, filtered, and concentrated under reduced pressure. The resulting residue was purified by column chromatography (\ce{SiO2}; hexane: \ce{CH2Cl2} 4:1) obtaining the product \textbf{5} (1.43 g, 88\%) as a yellow solid \cite{Lee2017}.

\ce{^1H} NMR (300 MHz, \ce{CDCl3}) $\delta$: 8.74-8.71 (d, \textit{J} = 8.9 Hz, 1H), 7.63-7.58 (m, 1H), 7.23-7.18 (dd, \textit{J} = 8.6, 6.6 Hz, 1H), 7.11-7.08 (d, \textit{J} = 8.8 Hz, 1H) ppm.

\subsection{Synthesis of compound 1}
\begin{figure}[hbt]
    \includegraphics[width=0.45\columnwidth]{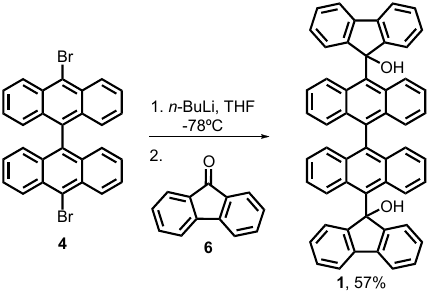}
\end{figure}
In a round bottom flask, compound \textbf{5} (500 mg, 0.98 mmol) and dry THF (25 mL) were added under argon atmosphere. The solution was cooled to -78 $^{\circ}$C and \textit{n}-BuLi solution (2.0 M in hexane, 1.1 mL, 2.15 mmol) was slowly added. The mixture was stirred for 1 h at -78 $^{\circ}$C and then 9H-fluoren-9-one (\textbf{6}, 350 mg, 1.90 mmol) in THF (5 mL) was added. The solution was slowly warmed to room temperature and stirred for 16h. Then water (20 mL) was added and extracted by \ce{CHCl3 }(3x50 mL). The organic layer was dried over anhydrous \ce{MgSO4}. The solvent was removed under reduced pressure and the residue was purified by column chromatography (\ce{SiO2}; hexane: \ce{CH2Cl2} 2:1) yielding the product \textbf{1} (260 mg, 57\%) as a yellow solid \cite{Zeng2012}.

\ce{^1H} NMR (300 MHz, \ce{CDCl3}) $\delta$: 9.91-9.88 (d, \textit{J} = 9.3 Hz, 2H), 7.92-7.89 (d, \textit{J} = 7.6 Hz, 4H), 7.55 7.45 (m, 10H), 7.32-7.26 (m, 4H), 7.20-7.18 (d, \textit{J} = 3.5 Hz, 4H), 7.15-7.12 (d, \textit{J} = 9.1 Hz, 2H), 7.02-6.99 (m, 2H), 6.88-6.83 (m, 2H), 6.76-6.70 (m, 2H), 2.80 (s, 2H) ppm.

\subsection{Synthesis of 4-(4-(methylthio)phenyl)-9H-fluoren-9-one (9)}
\begin{figure}[hbt]
    \includegraphics[width=0.5\columnwidth]{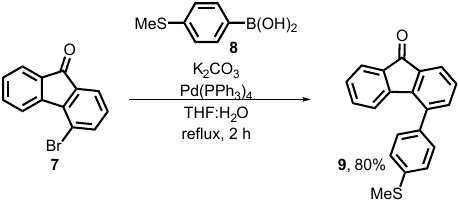}
\end{figure}
Over a deoxygenated mixture 4-bromo-9H-fluoren-9-one (\textbf{7}, 500 mg, 1.93 mmol), boronic acid \textbf{8} (486 mg, 2.90 mmol) and \ce{K2CO3} (800 mg, 5.79 mmol) in THF:\ce{H2O} (4:1, 50 mL) was added \ce{Pd(PPh3)4} (28 mg, 0.04 mmol). The resulting mixture was heated at reflux for 2 h. After cooling to room temperature, phases were separated and aqueous layer was extracted with EtOAc (3x 10 mL). Combined organic extracts were dried over anhydrous \ce{Na2SO4}, filtered and evaporated under reduced pressure. Crude product was purified by column chromatography (\ce{SiO2}; Hexanes:\ce{CH2Cl2} 4:1 to 1:1) affording \textbf{9} (460 mg, 80\%) as a yellow solid.

\ce{^1H} NMR (300 MHz, \ce{CDCl3}) $\delta$: 7.64 (q, \textit{J} = 5.0 Hz, 2H), 7.36 (s, 4H), 7.29 (d, \textit{J} = 4.7 Hz, 2H), 7.19 (dd, \textit{J} = 5.5, 3.2 Hz, 2H), 6.88 – 6.83 (m, 1H), 2.57 (s, 3H). ppm. \ce{^13C} NMR (75 MHz, \ce{CDCl3}) $\delta$: 193.8 (CO), 144.5 (C), 141.1 (C), 138.9 (C), 137.7 (C), 136.8 (CH), 136.0 (C), 134.8 (C), 134.4 (CH), 129.3 (CH), 128.8 (CH), 128.7 (CH), 126.4 (CH), 124.2 (CH), 123.3 (CH), 123.2 (CH), 15.6 (\ce{CH3}).ppm.HRMS (APCI): \ce{C20H15OS}; calculated: 303.0838, found: 303.0839.

\subsection{Synthesis of diol 10, precursor of SMe-2OS}
\begin{figure}[hbt]
    \includegraphics[width=0.52\columnwidth]{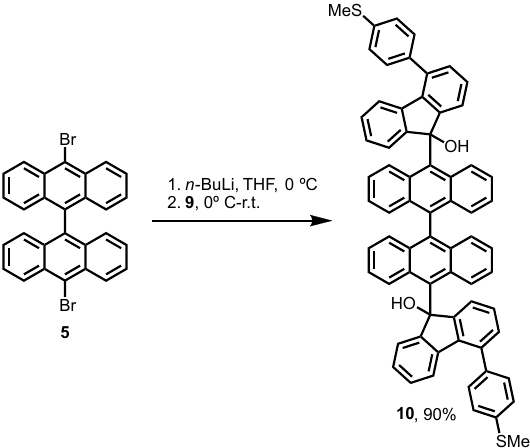}
\end{figure}
In a round bottom flask, compound \textbf{5} (77 mg, 0.15 mmol) and dry THF (5 mL) were added under argon atmosphere. The solution was cooled to 0 $^{\circ}$C and n-BuLi solution (2.5 M in hexane, 132 \textmu L, 0.33 mmol) was dropwise added. The mixture was stirred for 10 min at 0 $^{\circ}$C and then fluorenone \textbf{9} (100 mg, 0.38 mmol) in THF (1 mL) was added. The solution was slowly warmed to room temperature and stirred for 16h. Then water (10 mL) was added, and the mixture was extracted with EtOAc (3x5 mL). The organic layer was dried over anhydrous \ce{MgSO4} and filtered. The solvent was removed under reduced pressure and the residue was purified by column chromatography (\ce{SiO2}; Hexanes:\ce{CH2Cl2} 1:1 to 1:9) affording product \textbf{10} (120 mg, 90\%) as a greenish solid.

\ce{^1H} NMR (300 MHz, \ce{CDCl3}) $\delta$: 9.86 (d, \textit{J} = 9.3 Hz, 2H), 7.57 – 7.34 (m, 14H), 7.24 – 7.17 (m, 4H), 7.17 – 7.07 (m, 12H), 6.96 (d, \textit{J} = 8.6 Hz, 2H), 6.81 (t, \textit{J} = 7.2 Hz, 2H), 6.72 (t, \textit{J} = 7.8 Hz, 2H), 2.80 (s, 2H), 2.56 (s, 6H) ppm. \ce{^13C} NMR (300 MHz, \ce{CDCl3}) $\delta$ 153.4 (C), 152.9 (C), 138.8 (C), 138.4 (C), 137.4 (C), 135.8 (C), 135.7 (C), 134.1 (C), 132.7 (C), 132.0 (C), 131.5 (C), 131.4 (CH), 129.8 (CH), 129.1 (C), 128.9 (CH), 128.8 (CH), 128.7 (CH), 128.6 (CH), 128.5 (CH), 127.8 (CH), 127.3 (CH), 126.6 (CH), 126.3 (CH), 125.2 (CH), 125.0 (CH), 124.8 (CH), 124.6 (CH), 124.1 (CH), 124.0 (CH), 86.8 (C), 15.8 (\ce{CH3}) ppm. HRMS (APCI): \ce{C68H46O2S2}; calculated: 958.2939, found: 958.2944.

\newpage
\subsection{Spectroscopic data}

\begin{figure}[hbt]
    \includegraphics[width=0.9\columnwidth]{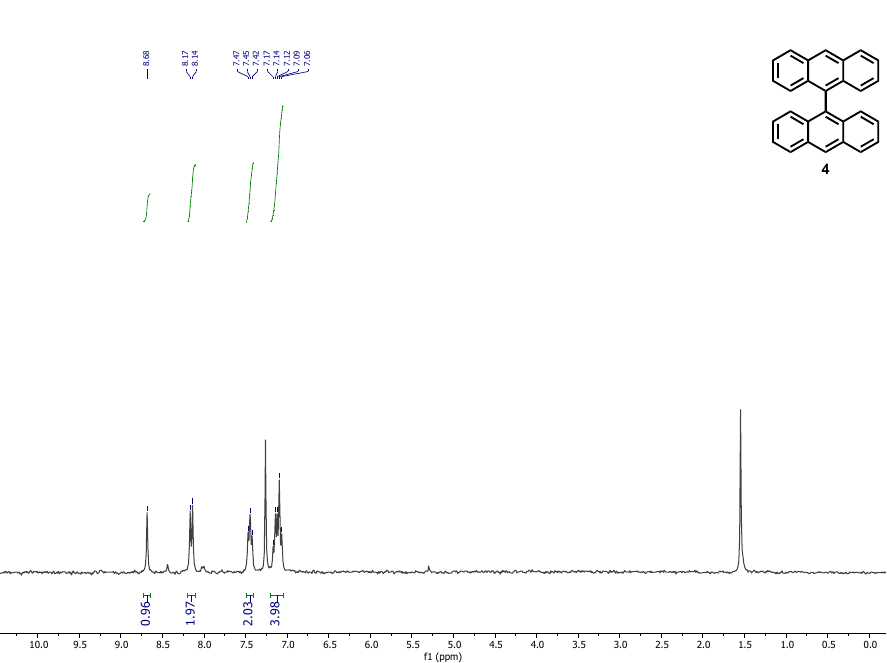}
     \caption{\ce{^1H} NMR (300 MHz, \ce{CDCl3}) spectrum of 9,9’-bianthracene (\textbf{4}).}
\end{figure}

\begin{figure}[hbt]
    \includegraphics[width=0.9\columnwidth]{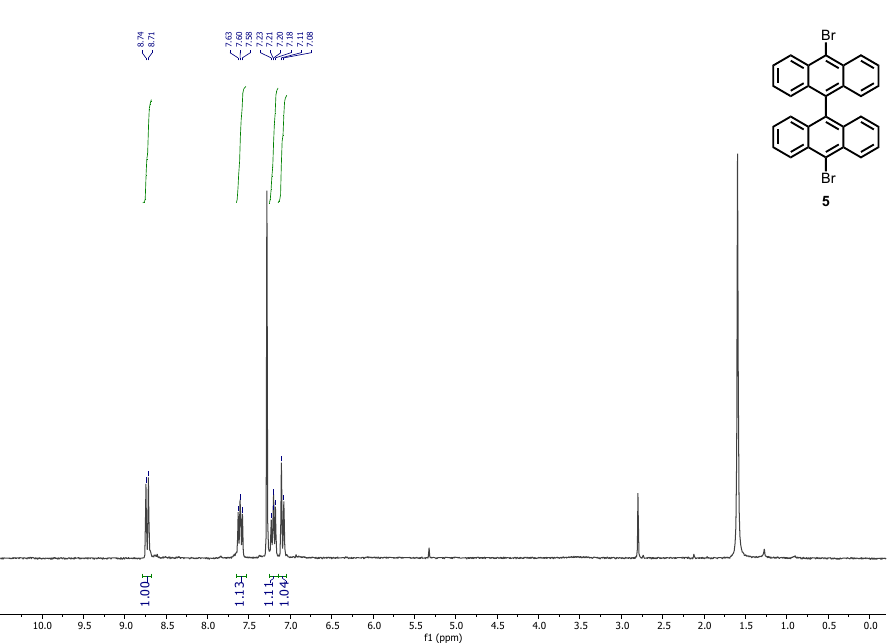}
    \caption{\ce{^1H} NMR (300 MHz, \ce{CDCl3}) spectrum of 10,10’-dibromo-9,9’-bianthracene (\textbf{5}).}
\end{figure}

\begin{figure}[hbt]
    \includegraphics[width=0.9\columnwidth]{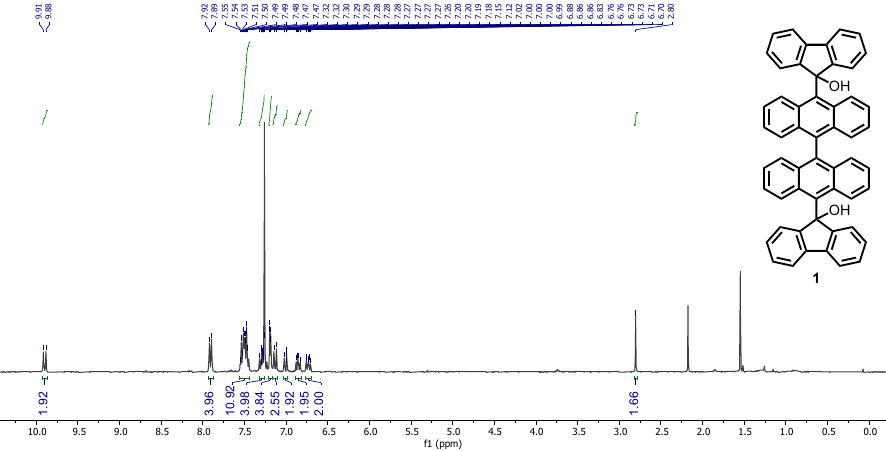}
    \caption{\ce{^1H} NMR (300 MHz, \ce{CDCl3}) spectrum of precursor \textbf{1}.}
\end{figure}

\begin{figure}[hbt]
    \includegraphics[width=0.9\columnwidth]{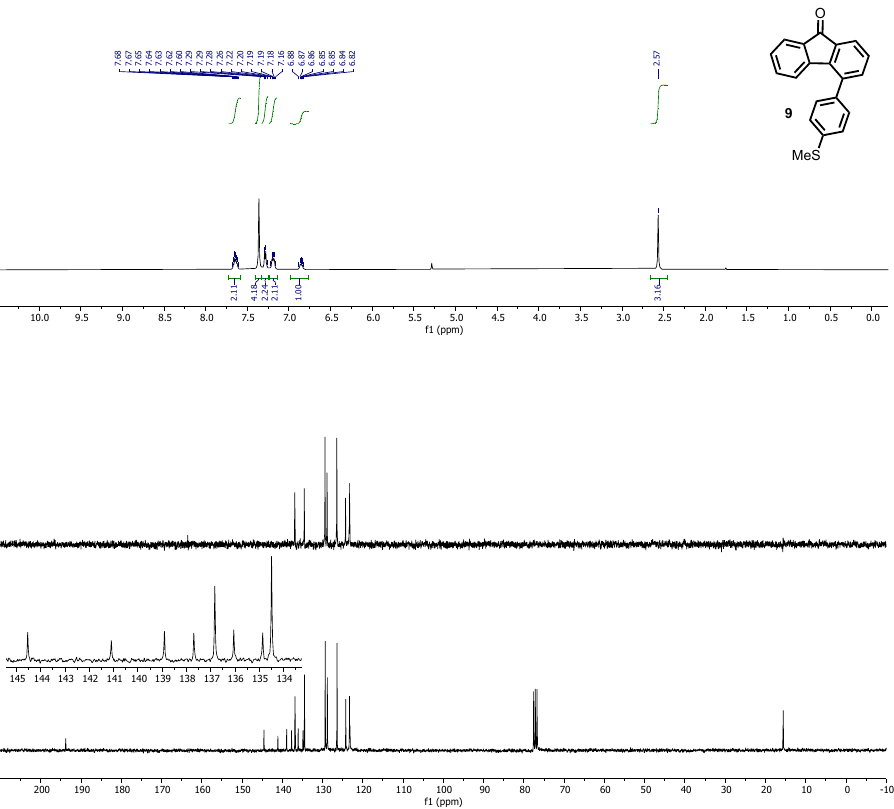}
    \caption{\ce{^1H} and \ce{^13C} NMR (300 MHz, \ce{CDCl3}) spectra of 4-(4-(methylthio)phenyl)-9H-fluoren-9-one (\textbf{9}).}
\end{figure}

\begin{figure}[hbt]
    \includegraphics[width=0.9\columnwidth]{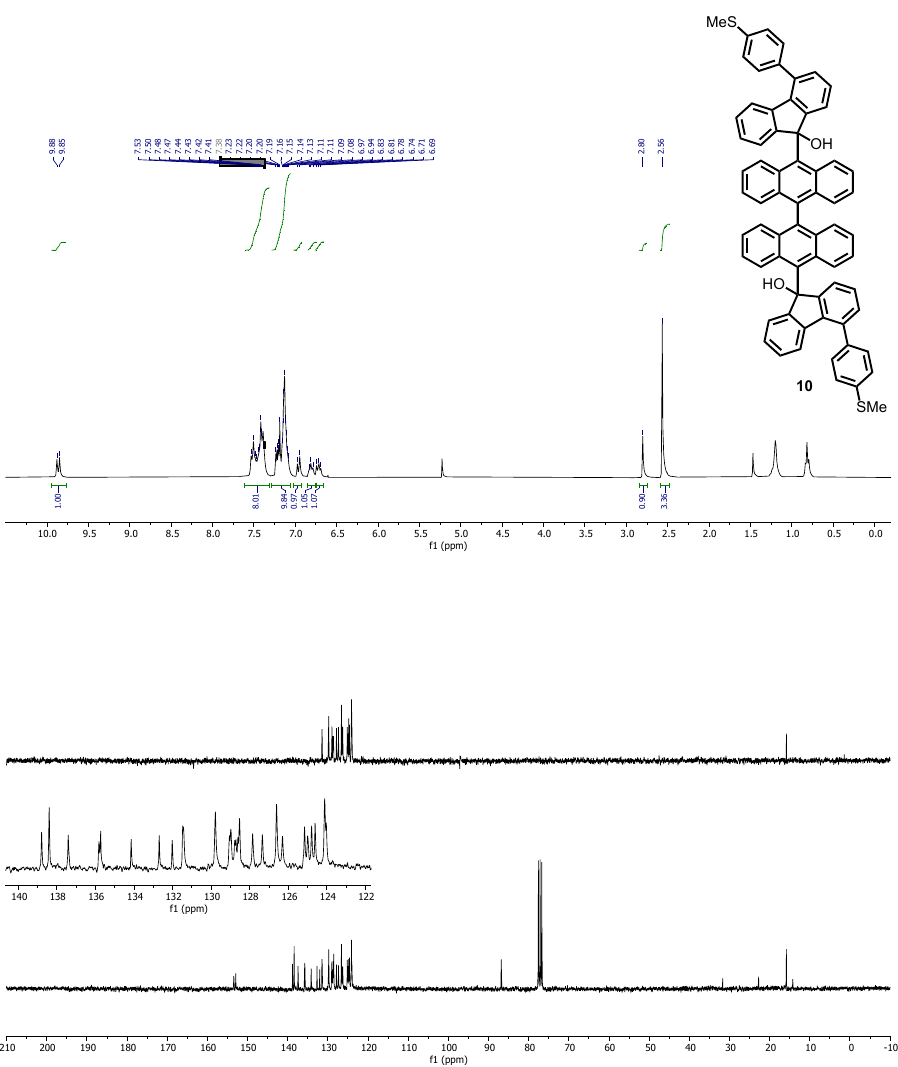}
    \caption{\ce{^1H} and \ce{^13C} NMR (300 MHz, \ce{CDCl3}) spectra of diol \textbf{10}.}
\end{figure}

\clearpage
\section{Complementary experimental data}

\begin{figure}[hbt]
    \includegraphics[width=\columnwidth]{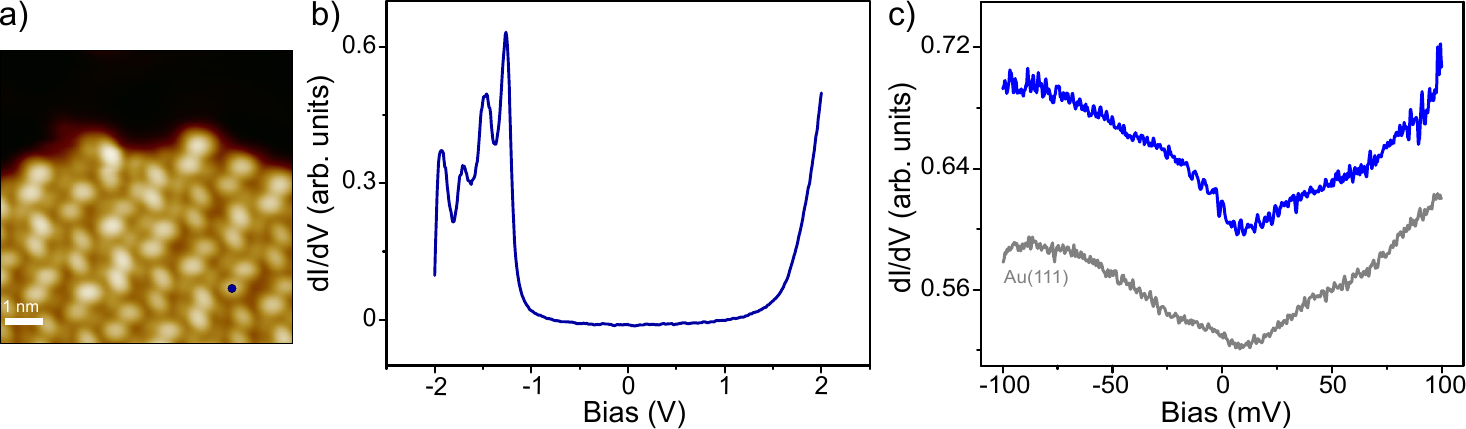}
\caption{Long- and short-range $dI/dV$ spectra measured on the molecular precursor (diol \textbf{1}) inside a close-packed island. (a) STM constant-current image of a close-packed domain formed by the molecular precursors deposited on the Au(111) surface ($V=1$ V; $I=30$ pA). (b) Long-range $dI/dV$ spectrum acquired in the position indicated by the blue circle in (a). Parameters: $V_{mod}=10$ mV, $I_{set}=500$ pA. (c) Short-range spectrum acquired on a molecule in the close-packed domain (blue curve), showing no magnetic fingerprints, due to the presence of passivating OH groups in diol \textbf{1}. Grey curve corresponds to a reference spectrum acquired on the bare Au(111) surface. Parameters: $V_{mod}=2$ mV, $I_{set}=400$ pA.
	}
 \label{fig:FigS1} 
\end{figure}

\begin{figure}[hbt]
    \includegraphics[width=0.9\columnwidth]{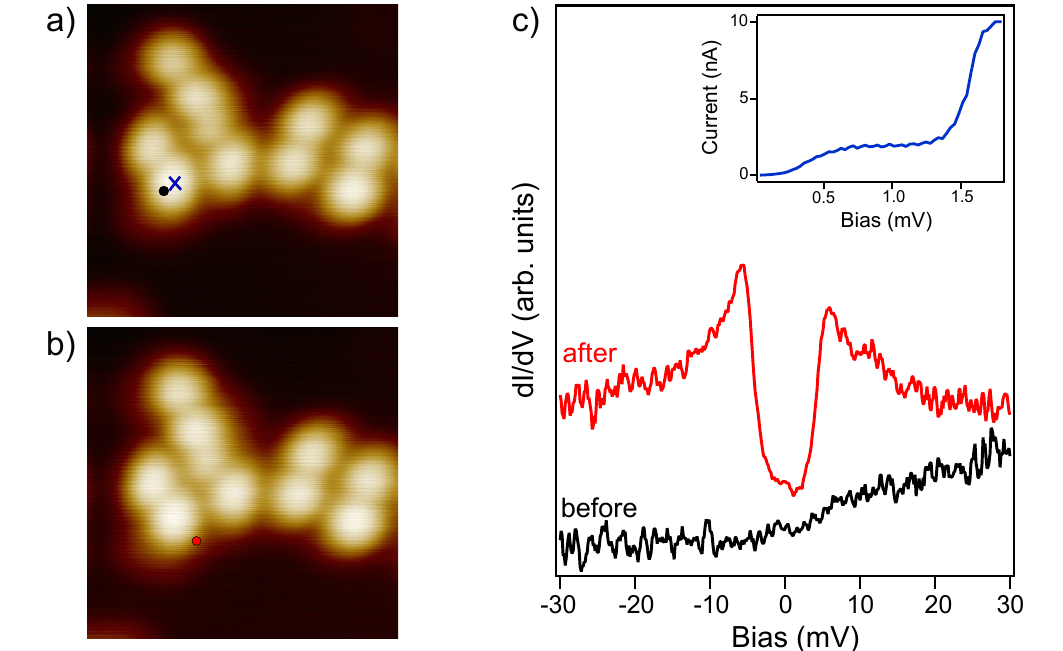}
\caption{On-surface generation of \textbf{2-OS} by electron-induced C-OH cleavage. (a) Constant-current STM image  of a small island of precursor molecules (diol \textbf{1} in Figure \ref{fig:precursors}). The blue cross indicates the position where the tip is stabilized (setpoint: $V=30$ mV, $I=10$ pA) before sweeping the bias. (b) Constant-current STM image of the same area after the bias sweep. (c) $dI/dV$ spectra measured before (black) and after (red) the bias sweep from 30 mV to 2 V at the positions indicated in (a,b) by the corresponding circles. The inset shows the variation of the tunneling current during the process. After the bias sweep, the diradical is generated, as clearly demonstrated by the appearance of the IET steps attributed to the two exchange-coupled spins. This experiment was performed at $T=1.3$ K. Topography parameters: $V=2$ V; $I=10$ pA. Spectroscopy parameters: $V_{mod}=0.8$ mV, $I_{set}=200$ pA.}
\label{fig:FigS2} 
\end{figure}	

\begin{figure}[hbt]
    \includegraphics[width=\columnwidth]{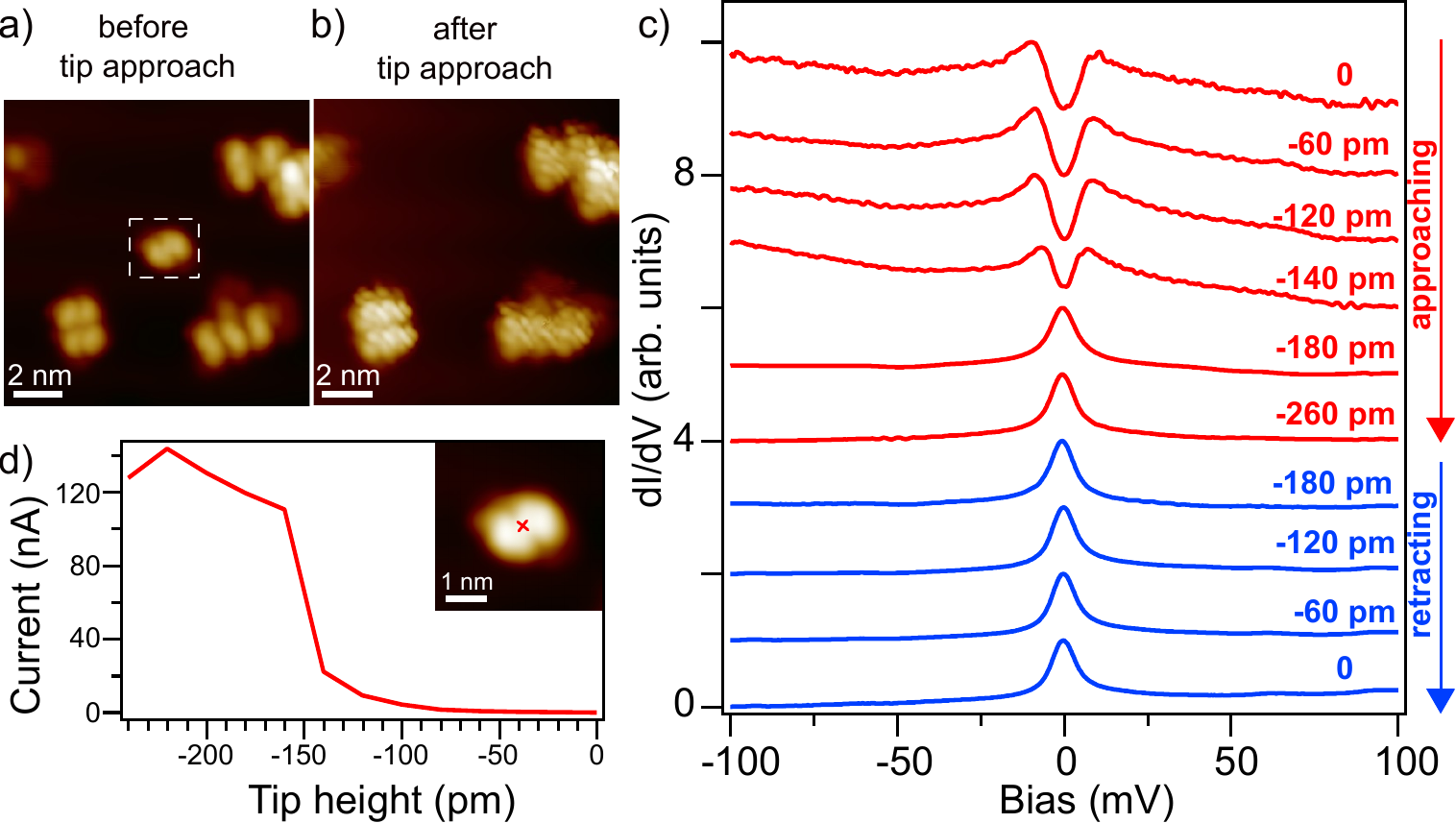}
\caption{Manipulation experiments on \textbf{2-OS}. (a) Constant-current STM image of the molecule on which the tip approach experiment is performed, as seen before starting the process ($V=-1.25$ V; $I=30$ pA); (b) Image of the same area after the manipulation, showing that the molecule has been unintentionally picked up by the tip. (c) Stack of normalized $dI/dV$ spectra recorded during the approach-retracting experiment on the \textbf{2-OS} molecule. The manipulation procedure is the following: we stabilize the tip above \textbf{2-OS}, in the position indicated by the cross in the inset in (d), at a starting height ($z=0$) determined by a setpoint of $V=100$ mV and $I=200$ pA; then we approach towards the molecule in steps of 20~pm, measuring $dI/dV$ spectra at each step, and finally rectract back to the starting point. Red curves are recorded during approaching, blue curves during retraction. We can observe the appearance of a Kondo peak and simultaneous disappearance of the IET steps at -180~pm. The Kondo feature persists while retracting the tip back to the starting position: this suggests that the change in the spectral feature is due to the quenching of one of the two radicals upon formation of a tip-molecule contact, as confirmed also by the image recorded at the end of the process (b), which shows that the molecule has been picked up during the process. (d) Current at $V=100$ mV recorded during the approach experiment presented in panel c). The sharp increase at -150~pm coincides with the change of magnetic fingerprint in the $dI/dV$ spectra. Spectroscopy parameters: $V_{mod}=2$ mV, $I_{set}=200$ pA.}
\label{fig:FigS3} 
\end{figure}

\begin{figure}[hbt]
    \includegraphics[width=0.7\columnwidth]{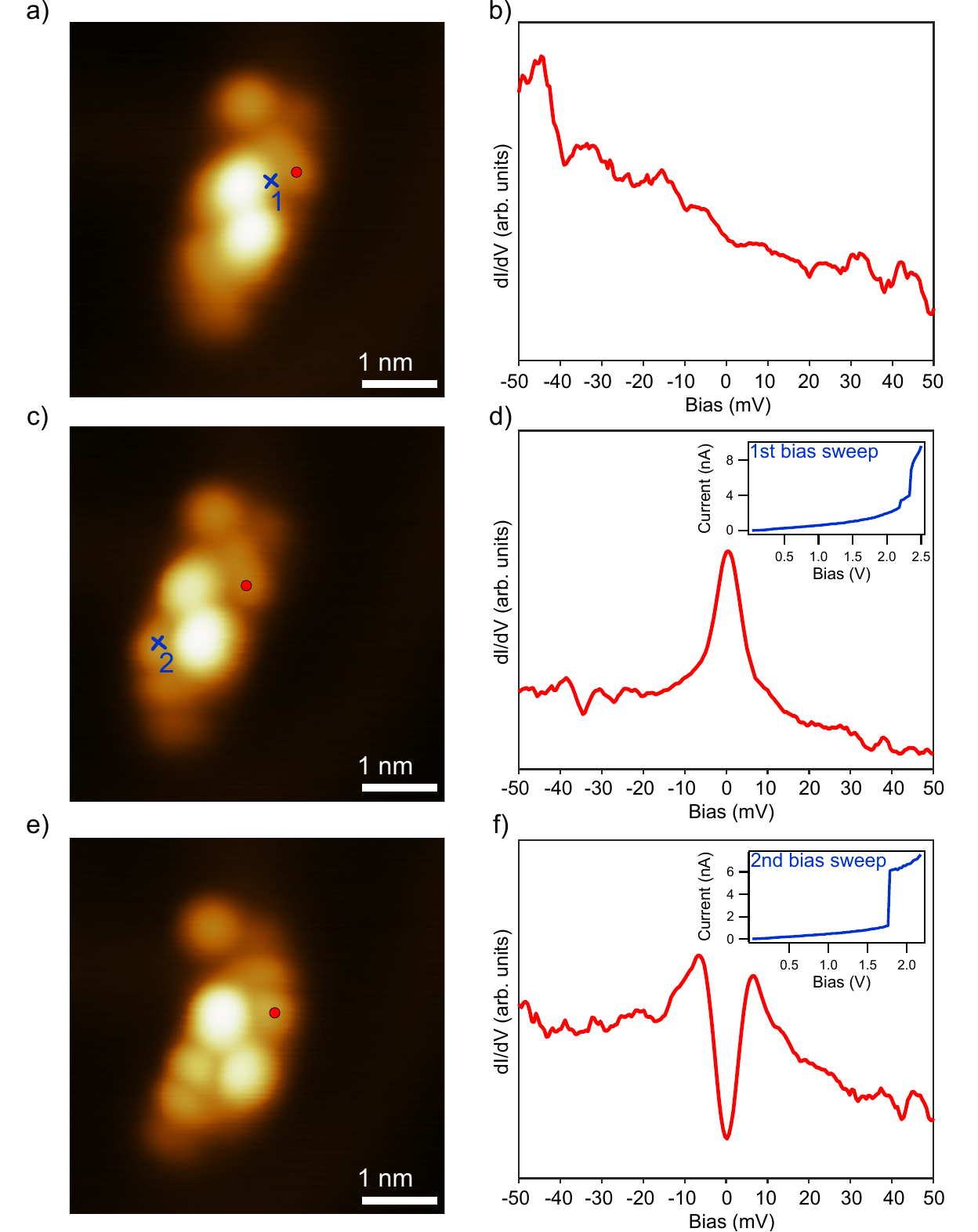}
\caption{On-surface generation of \textbf{SMe-2OS} by sequential electron-induced C-OH cleavage. (a) Constant-current STM image of the closed-shell precursor (diol \textbf{10} in Figure \ref{fig:precursors}) as deposited on the Au(111) surface. (b) Low bias $dI/dV$ spectrum measured at the position indicated by the red circle in (a) before the bias sweep, displaying no magnetic fingerprints, as expected for the closed-shell precursor. (c) Constant-current STM image of the molecule after the first bias sweep. The tip is stabilized in the position indicated by the blue cross in (a) (setpoint: $V=30$ mV, $I=10$ pA) before opening the feedback and raising the bias above 30 mV. (d) $dI/dV$ spectrum measured after the first bias sweep in the position indicated by the red circle in c), showing a Kondo peak that emerges as a consequence of the removal of one OH group. The inset reports the variation of the tunneling current during the bias sweep (from 30 mV to 2.5 V): the sudden jump at 2.3 V indicates the C-OH cleavage. (e) Constant-current STM image of the molecule after the second bias sweep, performed in the position indicated by the blue cross in (c), corresponding to the site where the second OH is located. (f) $dI/dV$ spectrum measured after the second bias sweep at the position marked by the red circle in e). The inset reports the variation of the current as a function of the bias (ramping from 30 mV to 2.1 V) with a steep increase at 1.8V. Now clear IET features appear, indicating the dissociation of the second OH group. Topography parameters: $V=1$ V; $I=30$ pA. Spectroscopy parameters: $V_{mod}=2$ mV, $I_{set}=500$ pA.}
\label{fig:FigS4} 
\end{figure}

\clearpage
\section{Complementary theoretical results}
\label{sec:theory}
\subsection{MF-Hubbard modeling based on Slater-Koster parametrization}
\label{sec:SK}
In order to explore the role of the torsion angles of \textbf{2-OS} we implemented an effective non-orthogonal tight-binding description based on a Slater-Koster parametrization \cite{SlKo.54.SimplifiedLCAOMethod}. Our model considers a single $p$ orbital per carbon site perpendicular to the anthracene (or fluorenyl) plane to which the atom belongs.

Consider two local $p_z$-orbitals $i$ and $j$, located arbitrarily at positions $\mathbf{R}_i$ and $\mathbf{R}_j$ and rotated by angles $\varphi_i$ and $\varphi_j$ (only) around the $x$ axis (oriented along the torsional bond of interest).
With $\mathbf{n}=(n_x, n_y, n_z)$ being the unit vector connecting the two orbital centers, and $d=|\mathbf{R}_i - \mathbf{R}_j|$ the distance between them, the effective hopping matrix element in the SK Hamiltonian connecting the basis orbitals can be expressed as
\begin{eqnarray}
t_{ij} &=& \left(t_{zz} \cos\varphi_i + t_{yz} \sin\varphi_i\right)\cos\varphi_j
+ \left(t_{yz} \cos\varphi_i + t_{yy} \sin\varphi_i\right)\sin \varphi_j
\end{eqnarray}
where
\begin{eqnarray}
t_{yz} &=& n_y n_z \left(V_{pp\sigma} - V_{pp\pi}\right)\\
t_{yy} &=& n_y^2 V_{pp\sigma} + (1-n_y^2) V_{pp\pi}\\
t_{zz} &=& n_z^2 V_{pp\sigma} + (1-n_z^2) V_{pp\pi}
\end{eqnarray}
are expressed in terms of the two-centre SK parameters
\begin{eqnarray}
V_{pp\alpha}(d) &=& V^0_{pp\alpha} \exp(-\beta_{pp\alpha} \frac{d - d_0}{d_0}), \qquad \alpha=\sigma, \pi.
\end{eqnarray}

\begin{figure}[h]
 \centering
    \includegraphics[width=0.7\textwidth]{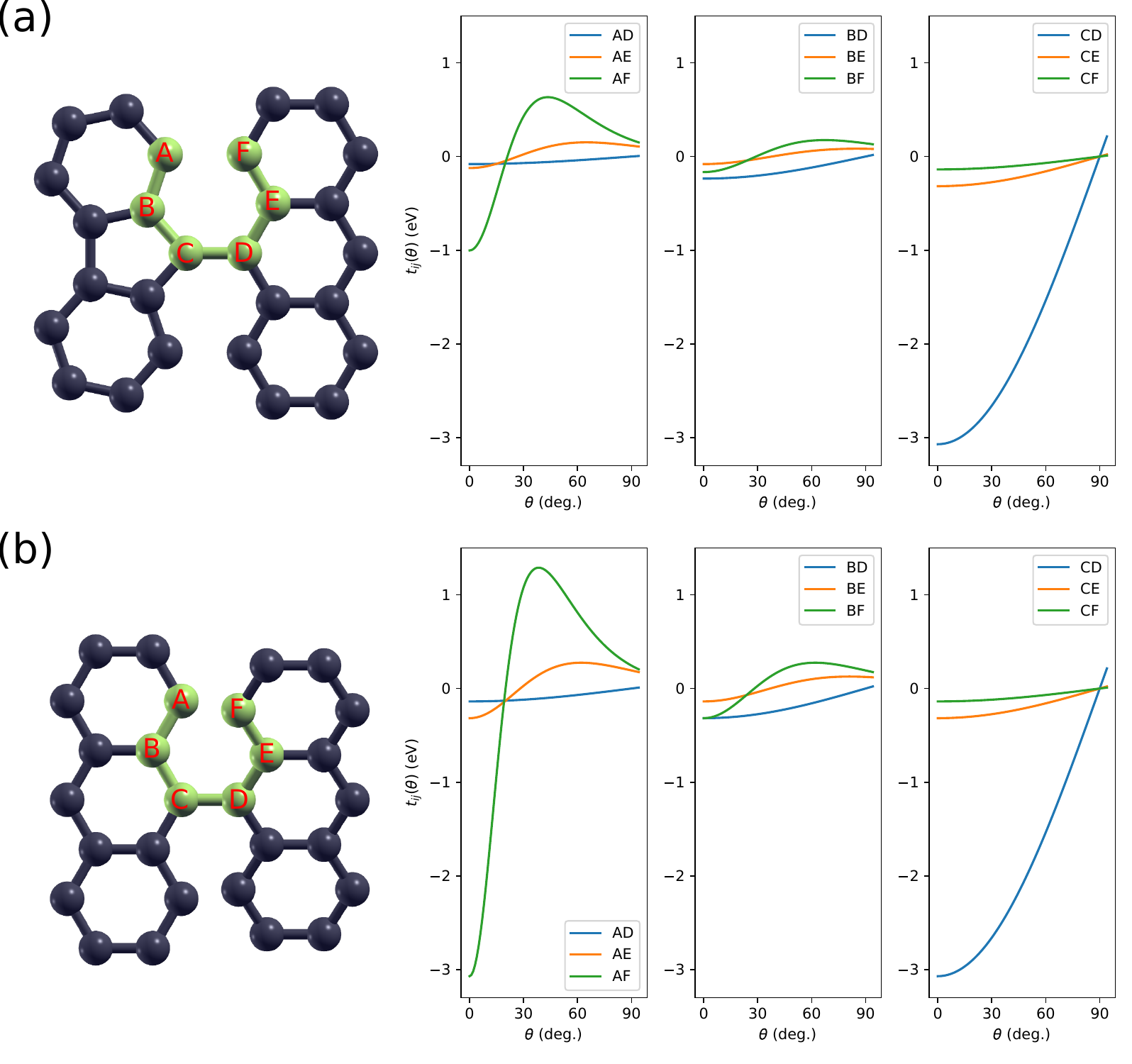}
\caption{Selective SK hopping matrix elements $t_{ij}$ between characteristic sites as a function of the torsional angle $\theta$ around the C--D bond connecting (a) fluorene and anthracene or (b) two anthracene units.}
\label{fig:SM-SK}
\end{figure}

\begin{figure}[h]
 \centering
    \includegraphics[width=0.7\textwidth]{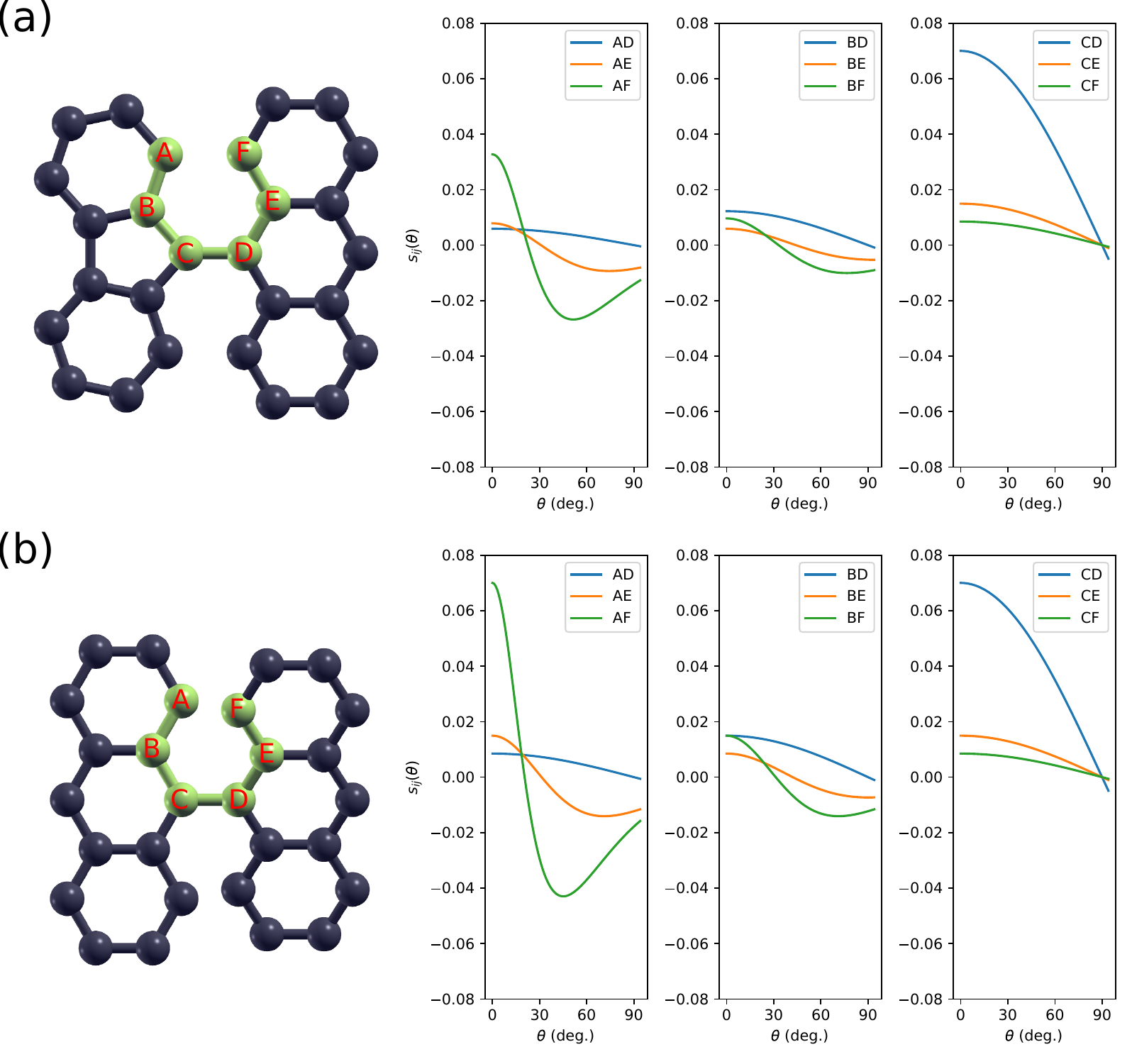}
\caption{Selective SK overlap matrix elements $s_{ij}$ between characteristic sites as a function of the torsional angle $\theta$ around the C--D bond
connecting (a) fluorene and anthracene or (b) two anthracene units.}
\label{fig:SM-SK2}
\end{figure}

Similarly, the orbital overlap can be expressed as
\begin{eqnarray}
s_{ij} &=& \left(s_{zz} \cos\varphi_i + s_{yz} \sin\varphi_i\right)\cos\varphi_j
+ \left(s_{yz} \cos\varphi_i + s_{yy} \sin\varphi_i\right)\sin \varphi_j
\end{eqnarray}
where
\begin{eqnarray}
s_{yz} &=& n_y n_z \left(S_{pp\sigma} - S_{pp\pi}\right)\\
s_{yy} &=& n_y^2 S_{pp\sigma} + (1-n_y^2) S_{pp\pi}\\
s_{zz} &=& n_z^2 S_{pp\sigma} + (1-n_z^2) S_{pp\pi}
\end{eqnarray}
and
\begin{eqnarray}
S_{pp\alpha}(d) &=& S^0_{pp\alpha} \exp(-\beta_{pp\alpha} \frac{d - d_0}{d_0}), \qquad \alpha=\sigma, \pi.
\end{eqnarray}
Following Ref.\cite{RePh.18.Modifiedspinorbit} we chose their parameters for our model: 
\begin{eqnarray}
V_{pp\sigma} &=& 6.20 \; \text{eV},\\
V_{pp\pi} &=& -3.07 \; \text{eV},\\
\beta_{pp\sigma} &=& 1.47, \\
\beta_{pp\pi} &=& 3.104,\\
S_{pp\sigma} &=& -0.140,\\
S_{pp\pi} &=& 0.07,\\
\beta'_{pp\sigma} &=& 0.77, \\
\beta'_{pp\pi} &=& 2.11,\\
d_0 &=& 1.42 \;\text{\AA}.
\end{eqnarray}
Using a custom implementation based on \textsc{sisl} \cite{zerothi_sisl}, some characteristic hopping matrix elements $t_{ij}$ between fluorenyl and anthracene sites are shown in Figure \ref{fig:SM-SK}. Similarly, the overlaps $s_{ij}$ are shown in Figure \ref{fig:SM-SK2}.

In addition to this effective one-orbital-per-site SK model, we consider in the main text (Figure 3) also a local onsite Coulomb repulsion $U$ in the mean-field approximation (MF-Hubbard). We solve iteratively for the symmetry-broken solutions corresponding to $S_z=0$ (AFM or unpolarized) and $S_z=1$ (FM)
using \textsc{hubbard} \cite{dipc_hubbard}.

In addition to this, we also checked the ground state of \textbf{2-OS} using CASSCF-NEVPT2 calculations, confirming the stability of the $S_z=0$ ground state, which is mostly degenerate with the $S_z=1$ excited state in gas phase.

\subsection{DFT study of surface adsorption geometry} \label{sec:DFTgeometry}
\begin{figure}[h]
 \centering
    \includegraphics[width=0.9\textwidth]{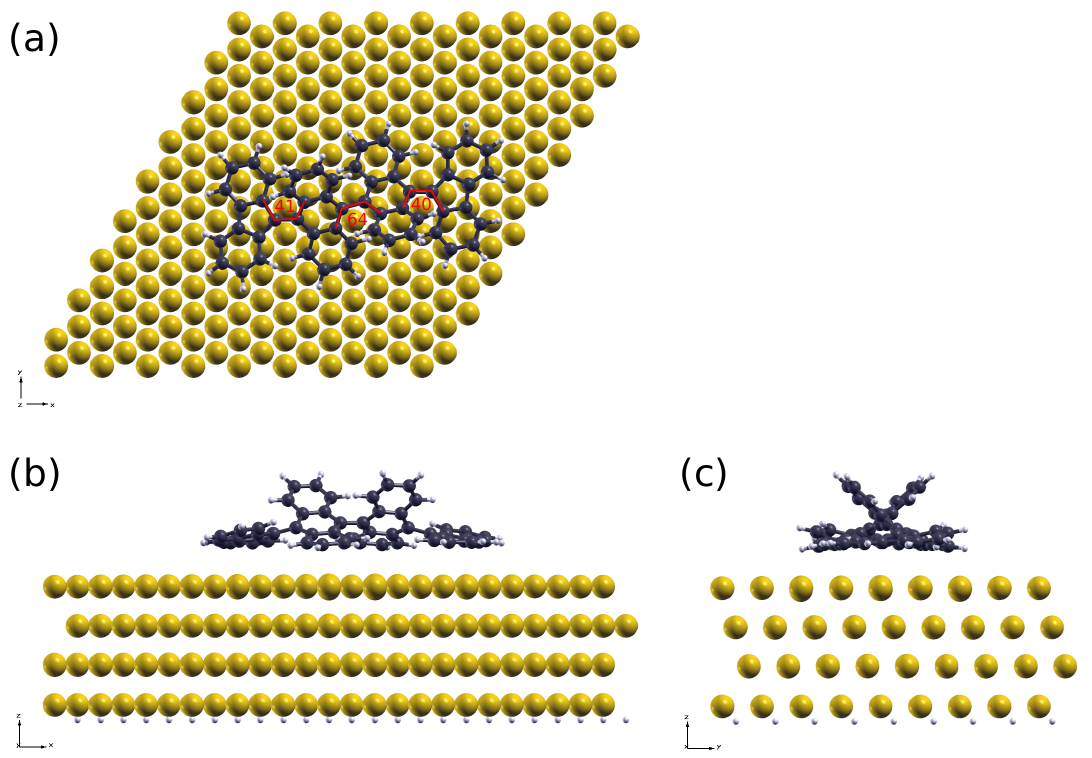}
\caption{Adsorption geometry of \textbf{2-OS} on Au(111) obtained with SIESTA. The three torsional angles indicated in panel (a) corresponds to $\theta_\mathrm{outer}^\mathrm{surf}=40^\circ$ and $\theta_\mathrm{inner}^\mathrm{surf}=64^\circ$. As reference, the corresponding angles in the gas phase were found to be $\theta_\mathrm{outer}^\mathrm{free}=54^\circ$ and $\theta_\mathrm{inner}^\mathrm{free}=71^\circ$.}
\label{fig:siesta-surface-geom}
\end{figure}

In our study, Density Functional Theory (DFT) calculations were executed using \textsc{Siesta} \cite{SoArGa.02.SIESTAmethodab} to understand the interactions between \textbf{2-OS} and the metallic surface. We applied the van der Waals density functional formulated by Dion \textit{et al.} \cite{DiRySc.04.VanderWaals} with the modified exchange by Klime{\v{s}}, Bowler and Michaelides \cite{KlBoMi.10.Chemicalaccuracyvan}, to effectively capture the dispersive forces at play. We used a double-$\zeta$ basis set for valence-electron wave function expansion, with orbital radii determined by a 100 meV energy shift and core electrons described via norm-conserving Troullier-Martins pseudopotentials. Brillouin zone sampling was conducted with a simple $1 \times 1 \times 1$ $k$-point grid. The Au(111) surface modeling utilized a $9 \times 9$-repeated, four-layer thick slab, characterized by a 4.08 Å lattice parameter, with an extended basis for the top atomic layer \cite{GaGaLo.09.Optimalstrictlylocalized} and hydrogen passivation on the bottom layer to mitigate spurious surface state interactions \cite{GoFeFr.08.FormationDispersiveHybrid}. The real-space grid was defined by a 300 Ry energy cutoff. Electronic occupations were smeared using a Fermi-Dirac distribution at a 300 K electronic temperature. Our self-consistency cycle criteria included convergence thresholds of less than $10^{-4}$ for the density and less than 1 meV for the Hamiltonian matrix elements, respectively. Geometry optimizations for the \textbf{2-OS} molecule and the top three layers of the Au(111) slab were performed using the conjugate gradient method, ensuring forces were below 0.01 eV/Å.
The adsorption geometry is shown in Fig.~\ref{fig:siesta-surface-geom}. In this adsorption configuration we could not stabilize the antiferromagnetic spin state of the gas-phase molecule, but it was recovered by increasing the molecule-surface distance by 0.5 \AA\ or more. The image shown in the inset of Figure 1c (main text) was obtained from the calculated charge density using the \textsc{WSxM} software \cite{Horcas2007}.

\clearpage
\subsection{Thermochemistry effects on DFT calculations}
\label{subsec:thermo}
In the following section we will address the effects of considering thermochemistry (vibrations, rotational motion, etc.) to DFT calculations on \textbf{2-OS}. 
As it is discussed in the main text, according to MF-Hubbard calculations the open-shell $S=0$ solution is lower in energy that the triplet. A variation of the angles $\theta_{o,i}$ affects the $\Delta E_{TS}$ energy separation, but it does not change the spin quantum number of the ground state. The effect of $\theta_{o,i}$ on the hoppings is calculated with a Slater-Koster parametrization in the case of MF-Hubbard, obtaining that $\Delta E_{TS}$ decreases when the angles are increased. 
The antiferromagnetic character of the kinetic exchange driven by hopping makes the $S=0$ solution more stable, becoming degenerate with a $S=1$ solution when two anthracene units, or an anthracene and an external fluorenyl unit, are perpendicular.

In Gaussian (g16)\cite{g16} it is possible to include the effects of temperature by the "\texttt{freq}" flag \cite{ochterski2000thermochemistry}. This flag entails the computation of different contributions to the entropy ($S_{tot}$) and internal thermal energy ($E_{tot}$). Specifically, this program considers the contributions from translation, electronic motion, rotational motion and vibrational motion. Then, it considers a thermal correction to the enthalpy and free energy, respectively, as:
\begin{equation}
H_{corr} = E_{tot} + k_BT,
\end{equation}    
and
\begin{equation}
G_{corr} = H_{corr} - TS_{tot},
\end{equation}
where $S_{tot} = S_t + S_r + S_v + S_e$ and $E_{tot}= E_t+E_r+E_v+E_e$ consider the different contributions that have been mentioned, where $t$, $r$, $v$ and $e$ refer, respectively, to the contributions from translation, rotational motion, vibrational motion and electronic motion.

\begin{figure}[h]
 \centering
    \includegraphics[width=0.4\textwidth]{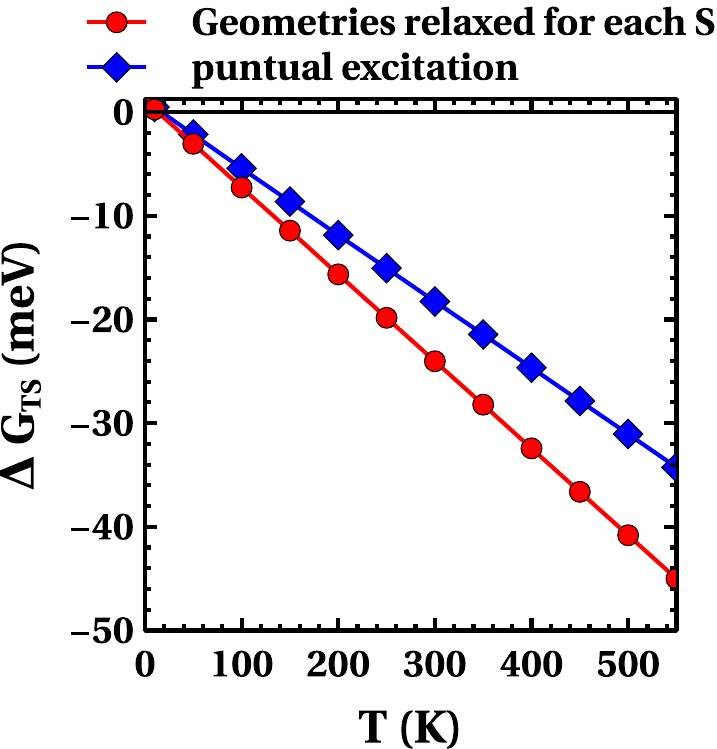}
\caption{Variation of $G_T - G_S = \Delta G_{TS}$ with the temperature ($T$), calculated with DFT. It was used the PBE exchange-correlation density functional \cite{PeBuEr.96.Generalizedgradientapproximation} and the STO-3G basis \cite{hehre1969self,collins1976self} with the Gaussian package\cite{g16} and the thermochemistry flag. For the red circles the geometry was relaxed for both spin solutions, and for the blue squares it was used the $S=1$ relaxed geometry for all the calculations. The geometries were relaxed without the "\texttt{freq}" flag.}
\label{fig:SMthermo}
\end{figure}

By comparing the sum of electronic and thermal free energies for the $S=1$ and open-shell $S=0$ solutions ($G_{S,T}$) for different values of the temperature with a relaxed gas phase geometry ($\theta_i\approx 90^\circ$), we can conclude that the thermochemistry effects are responsible of the $S=1$ ground state in gas phase (see Figure \ref{fig:SMthermo}), since for low $T$ the $S=0$ is the ground state but very soon the $S=1$ becomes more stable when the temperature is increased. Then, if the geometry is planarized, the $S=0$ solution is stabilized in account of an enhancement of the kinetic antiferromagnetic exchange, in accordance with MF-Hubbard.

\begin{table}[htp]
 \begin{center}
\begin{tabular}{| l | c | c | c | c | c |}
\hline
     & PBE/STO-3G (5K) & PBE/STO-3G (100K) & BLYP/6-31G (5K) & BLYP/6-31G (100K)  \\ \hline
    $\Delta G_{TS}$ (meV) & $0.7$ & $-7$ & $-0.2$ & $-9$ \\
\hline
\end{tabular}
\end{center}
\caption{Difference in energy between the open-shell $S=0$ and $S=1$ solutions for different density functional/basis and temperatures. The geometries were relaxed for every $S$ in gas phase and without thermochemistry.}\label{table1}
 \end{table}

\clearpage
\section{Bibliography}

%